\documentclass[sn-mathphys,Numbered,pdflatex]{sn-jnl}

\usepackage{graphicx}%
\usepackage{multirow}%
\usepackage{amsmath,amssymb,amsfonts}%
\usepackage{mathrsfs}%
\usepackage{bm}%
\usepackage{xcolor}%
\usepackage{textcomp}%
\usepackage{manyfoot}%
\usepackage{booktabs}%
\usepackage{algorithm}%
\usepackage{algorithmicx}%
\usepackage{algpseudocode}%

\providecommand{\onlinecite}[1]{\cite{#1}}
\makeatletter
\@ifundefined{ruledtabular}{\newenvironment{ruledtabular}{}{}}{}
\makeatother

\begin{document}

\title[Entropy-maximized active learning for MLIPs]{Dataset-aware entropy-maximized active learning for machine-learned interatomic potentials}

\author{\fnm{Meiyan} \sur{Wang}}

\author{\fnm{Rishi} \sur{Rao}}

\author*{\fnm{Li} \sur{Zhu}}\email{li.zhu@rutgers.edu}

\affil{\orgdiv{Department of Physics}, \orgname{Rutgers University}, \orgaddress{\city{Newark}, \state{New Jersey} \postcode{07102}, \country{USA}}}

\abstract{
We present an active learning framework for efficiently generating training data for machine-learned interatomic potentials (MLIPs).
The method combines local entropy-driven molecular dynamics with global dataset-aware filtering: a per-configuration entropy term biases MD trajectories toward structurally diverse snapshots, while a global entropy measure, the log-determinant of the fingerprint covariance matrix of the entire dataset, selects only those configurations that provide genuinely new information.
We employ dual covariance modes (per-atom for disordered structures and per-config for ordered phases) to achieve broad coverage of configuration space.
Combined with a pre-trained foundation model (Allegro-OAM-L) and analytical fingerprint gradients from Gaussian overlap matrix eigenvalues, the framework produces high-quality domain-specific potentials with near- or sub-meV/atom accuracy on test data drawn from the same distribution at training-set sizes of order $10^{2}$ to $10^{3}$ entropy-selected DFT-labeled structures.
We demonstrate the method on three systems spanning diverse bonding types and pressure-driven phase transitions: carbon (covalent), silicon (covalent/metallic), and NaCl (ionic).
In learning curve comparisons against random molecular dynamics sampling at matched training set sizes ($N = 100$ to $800$), evaluated over three independent training-set draws per condition, entropy-driven sampling achieves a factor of approximately $3$ to $10$ lower energy MAE at $N = 800$ on in-distribution holdouts across the three systems, with the magnitude of the gain depending on the bonding type and the size at which the random-MD baseline saturates.
On a curated independent test set of equilibrium and thermal-MD configurations spanning each system's relevant phases and pressures, the entropy-driven energy-MAE advantage persists for all three systems, while force and stress accuracy is more test-distribution dependent: entropy-driven sampling matches or improves on random for silicon and NaCl, but underperforms random on carbon force and stress, a tradeoff that is consistent with the broader fingerprint-space coverage of the entropy pool over-representing distorted configurations relative to the near-equilibrium structures emphasized by the curated test.
For carbon, the reversal is resolved at no additional DFT cost by an $80/20$ entropy-plus-anchor pool (80\% entropy-driven + 20\% near-equilibrium random-MD snapshots, $\max|\mathbf{F}|<2$~eV/\AA, at fixed $N$), which matches pure-entropy energy accuracy on the independent test while recovering the random-pool force and stress accuracy.
}


\maketitle

\section{Introduction}
\label{sec:intro}

Machine-learned interatomic potentials (MLIPs) have emerged as a powerful approach to atomistic simulation, achieving near-ab initio accuracy at a fraction of the computational cost~\cite{behler2016perspective,deringer2019machine,unke2021machine}.
Modern MLIPs span a spectrum of architectures and representations: high-dimensional neural networks built on atom-centered symmetry functions~\cite{behler2007generalized}, Gaussian approximation potentials~\cite{bartok2010gap} with smooth-overlap-of-atomic-positions~\cite{bartok2013soap} or atomic-cluster-expansion~\cite{drautz2019ace} descriptors, deep-learning packages such as SchNet~\cite{schutt2018schnet} and DeePMD-kit~\cite{wang2018deepmd}, and message-passing equivariant neural networks including NequIP~\cite{batzner2022nequip}, MACE~\cite{batatia2022mace}, and Allegro~\cite{musaelian2023learning} have collectively established the modern accuracy and efficiency trade-off for element-specific models.
Recent universal foundation models, including MACE-MP-0~\cite{batatia2024foundation}, CHGNet~\cite{deng2023chgnet}, MatterSim~\cite{yang2024mattersim}, GNoME~\cite{merchant2023gnome}, and Allegro-OAM-L~\cite{fu2025allegro}, have demonstrated impressive transferability across the periodic table.
However, for quantitative studies of specific materials under targeted conditions, particularly at high pressures where training data are sparse, fine-tuning on domain-specific density functional theory (DFT) data remains essential.

The central challenge in MLIP training is generating a diverse and compact training set.
Random sampling of molecular dynamics (MD) trajectories produces highly correlated structures, requiring thousands of DFT calculations to adequately cover configuration space.
Entropy-maximized sampling~\cite{karabin2020entropy} addresses this by adding a diversity-driving term to the potential energy surface, but the original per-configuration formulation suffers from self-averaging: independently generated configurations are individually diverse but collectively redundant.

Several active learning strategies have been developed for MLIPs.
Concurrent learning frameworks such as DP-GEN~\cite{zhang2020dpgen} use committee disagreement among an ensemble of models to identify uncertain configurations during MD exploration, triggering on-the-fly DFT labeling and retraining; committee-based neural network potentials more broadly~\cite{schran2020committee} apply the same disagreement principle to control generalization error during simulation.
Bayesian Gaussian-process methods such as FLARE~\cite{vandermause2020flare} exploit the closed-form predictive variance of the surrogate to flag and label uncertain configurations during MD, and de novo exploration protocols using configuration-averaged kernel metrics~\cite{bernstein2019denovo} bootstrap training datasets without prior reference configurations.
Uncertainty-driven dynamics (UDD)~\cite{nebgen2023udd} directly modifies the potential energy surface to bias sampling toward high-uncertainty regions.
D-optimal active learning~\cite{podryabinkin2017active,shapeev2017moment} selects training points that maximize the information matrix determinant in the space of model features, minimizing parameter uncertainty.
These approaches share a common requirement: they depend on model-specific uncertainty estimates, such as ensemble variance, Bayesian posteriors, or feature-space leverage scores, that must be recomputed or retrained as the model evolves.

In this work, we introduce a dataset-aware active learning framework whose selection criterion is decoupled from the model architecture.
Our global entropy filter, the log-determinant of the fingerprint covariance matrix computed over the entire accumulated dataset, is a D-optimal-like objective~\cite{podryabinkin2017active} formulated in a fixed structural descriptor space (GOM fingerprints~\cite{zhu2016fingerprint}), making it compatible with any black-box potential including pre-trained foundation models.
Local entropy-biased MD~\cite{karabin2020entropy} generates candidate structures with high internal diversity, while the global filter accepts only those that expand the dataset's information content, resolving the self-averaging problem of purely local entropy approaches~\cite{subramanyam2025}.
Our approach combines four key ingredients: (i)~structural fingerprints with analytical gradients enabling force-based entropy-biased MD, (ii)~local entropy maximization for efficient exploration, (iii)~global dataset-aware covariance tracking with dual per-atom and per-config modes for selection, and (iv)~a pre-trained foundation model (Allegro-OAM-L~\cite{fu2025allegro,barroso2024omat24}) providing physically reasonable forces throughout the sampling process.
The exploration phase still depends on the base potential, but the dataset-selection criterion does not require ensemble training, committee disagreement, or model-specific uncertainty estimates at any stage of the pipeline.

We validate the framework on three chemically distinct systems: carbon at 0 to 100~GPa (covalent, with van der Waals interlayer interactions), silicon at 0 to 20~GPa (covalent/metallic transition), and NaCl at 0 to 40~GPa (ionic with B1$\to$B2 structural transition).
In each case, learning curve comparisons against random MD sampling, evaluated over three independent training-set draws per condition, demonstrate a factor of approximately $3$ to $10$ data efficiency at $N = 800$ training structures, with system-dependent magnitude.

\section{Method}
\label{sec:method}

\subsection{Structural fingerprints}
\label{sec:fingerprints}

We employ the GOM fingerprints of Ref.~\onlinecite{zhu2016fingerprint}.
For each atom~$i$ in a periodic cell, the fingerprint is constructed by (1)~collecting all neighbors~$j$ within a cutoff radius~$r_c$, (2)~building the overlap matrix
\begin{equation}
O_{ab} = \left(\frac{4\pi\alpha_a\alpha_b}{\alpha_a + \alpha_b}\right)^{3/2}
\exp\!\left(-\frac{r_{ab}^2\,\alpha_a\alpha_b}{\alpha_a + \alpha_b}\right),
\label{eq:overlap}
\end{equation}
where $\alpha_a = 1/(2r_{\mathrm{cov},a}^2)$ depends on the covalent radius, (3)~applying a smooth damping function $f(r) = (1 - r^2/r_c^2)^{N_c-1}$ to ensure continuity at the cutoff, and (4)~diagonalizing the damped overlap matrix to obtain eigenvalues $\{w_m\}$ in descending order, forming the fingerprint vector $\bm{q}_i \in \mathbb{R}^d$.
The dimension $d$ is set by the \texttt{natx} parameter: the $d$ largest eigenvalues are retained (zero-padded if fewer than $d$ neighbors lie within $r_c$), and eigenvalues are not normalized, preserving the absolute scale of the overlap integrals.
For multi-component systems, separate overlap matrices are constructed for each chemical species pair, and their eigenvalue spectra are concatenated.

A crucial advantage of GOM fingerprints is the availability of analytical gradients via the Hellmann-Feynman theorem:
\begin{equation}
\frac{\partial q_{i,m}}{\partial r_{j,k}} = \langle v_m | \frac{\partial O}{\partial r_{j,k}} | v_m \rangle,
\label{eq:fp_grad}
\end{equation}
where $|v_m\rangle$ is the eigenvector corresponding to eigenvalue $w_m$.
Strain derivatives $\partial q_{i,m}/\partial\varepsilon_{\alpha\beta}$ are similarly available, enabling stress computations.

\subsection{Global entropy measure}
\label{sec:entropy}

Given a dataset $\mathcal{D}$ of fingerprints collected from previously accepted configurations, we define the global entropy as
\begin{equation}
H(\mathcal{D}) = \log\det\bm{\Sigma}(\mathcal{D}),
\label{eq:entropy}
\end{equation}
where the covariance matrix is
\begin{equation}
\bm{\Sigma} = \frac{1}{N}\sum_{i\in\mathcal{D}} \bm{q}_i\bm{q}_i^T
- \bm{\mu}\bm{\mu}^T + \lambda\,\bm{I},
\label{eq:covariance}
\end{equation}
with $\bm{\mu} = N^{-1}\sum_i \bm{q}_i$ the dataset mean and $\lambda$ a regularization parameter.
Here $N$ denotes the total number of fingerprint vectors in the dataset: for per-atom mode, $N$ is the cumulative number of individual atomic fingerprints across all accepted configurations; for per-config mode, $N$ is the number of accepted configurations.
We use the biased (maximum-likelihood) covariance estimator, which is appropriate since $N \gg d$ in practice.
Equation~(\ref{eq:entropy}) corresponds to the differential entropy of a Gaussian distribution fitted to the fingerprint distribution and is maximized when the data spans a large, uncorrelated volume in fingerprint space.
This objective is closely related to D-optimal experimental design~\cite{podryabinkin2017active}, which maximizes $\log\det(\bm{X}^T\bm{X})$ of the design matrix to minimize parameter uncertainty.
The key difference is that D-optimality is typically formulated in model feature space and thus depends on the model architecture, whereas our criterion operates in a fixed structural descriptor space (GOM eigenvalues), making it agnostic to the choice of MLIP and compatible with black-box foundation models.

This global measure contrasts with the per-configuration entropy of Ref.~\onlinecite{karabin2020entropy},
\begin{equation}
S_{\mathrm{local}} = \frac{1}{N_{\mathrm{at}}} \sum_i \log(N_{\mathrm{at}} \cdot \delta q_{\min,i}),
\label{eq:local_entropy}
\end{equation}
where $N_{\mathrm{at}}$ is the number of atoms in the configuration and $\delta q_{\min,i} = \min_{j\neq i}\|\bm{q}_i - \bm{q}_j\|_2$ is the Euclidean nearest-neighbor distance, in (raw, unnormalized) fingerprint space, between atom $i$ and any other atom in the same configuration.
The product $N_{\mathrm{at}}\cdot\delta q_{\min,i}$ is dimensionless: $\delta q_{\min,i}$ has units of (eigenvalue scale) and $N_{\mathrm{at}}$ acts as a density-of-states correction, so the logarithm is well-defined in the $N_{\mathrm{at}}\to\infty$ limit~\cite{karabin2020entropy}.
$S_{\mathrm{local}}$ measures diversity within a single configuration only, and consequently leads to self-averaging: independently generated configurations are individually diverse but collectively redundant.

\subsubsection{Incremental statistics}

Rather than storing all fingerprints, we maintain sufficient statistics $(\sum\bm{q},\;\sum\bm{q}\bm{q}^T,\;N)$, requiring $O(d^2)$ memory regardless of dataset size.
The cost of adding a new configuration is independent of the accumulated dataset size and scales as $O(d^2)$ in per-config mode and $O(N_{\mathrm{at}}\,d^2)$ in per-atom mode.
The marginal information gain $\Delta H$ in Eq.~(\ref{eq:acceptance}) is evaluated using the matrix-determinant lemma, avoiding explicit recomputation of the determinant after every candidate.

\subsubsection{Dual-mode covariance}

We maintain two separate datasets:
\begin{itemize}
\item \textbf{Per-atom mode}: Individual atomic fingerprints are added. The covariance tracks diversity of local atomic environments, producing disordered structures.
\item \textbf{Per-config mode}: The configuration-averaged fingerprint $\bar{\bm{q}} = N_{\mathrm{at}}^{-1}\sum_i \bm{q}_i$ is added. The covariance tracks diversity of bulk-averaged structural character, preserving crystalline order.
\end{itemize}
Each MD run is randomly assigned to one mode (default: 60\% per-atom, 40\% per-config), ensuring coverage of both ordered and disordered regions of configuration space.

\subsection{Modified potential energy surfaces}
\label{sec:pes}

Our framework uses two distinct modified potential energy surfaces for exploration and selection, respectively.

\subsubsection{Exploration: local entropy-biased MD}

During the exploration phase, MD trajectories are biased using the local per-configuration entropy [Eq.~(\ref{eq:local_entropy})]:
\begin{equation}
E_{\mathrm{explore}} = E_{\mathrm{base}}(\bm{r}) - k \cdot S_{\mathrm{local}}(\bm{r}),
\label{eq:explore_energy}
\end{equation}
where $E_{\mathrm{base}}$ is provided by the pre-trained Allegro-OAM-L model~\cite{fu2025allegro} (or a fine-tuned version thereof) and $k$ controls the entropy weight.
The local entropy $S_{\mathrm{local}}$ depends only on the nearest-neighbor fingerprint distances within the current configuration, making it computationally inexpensive (no dataset bookkeeping during MD) and providing smooth, well-defined gradients for the integrator.

\subsubsection{Selection: global dataset entropy}

Independently of the exploration energy, the global log-determinant entropy [Eq.~(\ref{eq:entropy})] is used as a post-hoc selection criterion (Section~\ref{sec:generation}).
The global entropy gain $\Delta H$ [Eq.~(\ref{eq:acceptance})] is evaluated for each candidate snapshot collected during MD but does not enter the equations of motion.

\subsubsection{Refinement: global entropy optimization}

In the optional refinement phase (Section~\ref{sec:generation}), near-threshold candidates are optimized on the global entropy surface:
\begin{equation}
E_{\mathrm{refine}} = E_{\mathrm{base}}(\bm{r}) - k_{\mathrm{refine}} \cdot \log\det\bm{\Sigma}(\mathcal{D} \cup \{\mathrm{current}\}),
\label{eq:refine_energy}
\end{equation}
where the forces are
\begin{equation}
F_{i,k}^{\mathrm{ent}} = \frac{2}{N}\sum_j \bm{w}_j^T \frac{\partial\bm{q}_j}{\partial r_{i,k}},
\label{eq:force_peratom}
\end{equation}
with the per-atom weight vector $\bm{w}_j = \bm{\Sigma}^{-1}(\bm{q}_j - \bm{\mu})$ in per-atom mode (the sum over $j$ ranges over atoms in the current configuration), and
\begin{equation}
F_{i,k}^{\mathrm{ent}} = \frac{2}{N\cdot N_{\mathrm{at}}} \bm{w}_{\bar{q}}^T \sum_j \frac{\partial\bm{q}_j}{\partial r_{i,k}},
\label{eq:force_perconfig}
\end{equation}
with the per-configuration weight vector $\bm{w}_{\bar{q}} = \bm{\Sigma}^{-1}(\bar{\bm{q}} - \bm{\mu})$ in per-config mode, where $\bar{\bm{q}} = N_{\mathrm{at}}^{-1}\sum_j \bm{q}_j$ is the configuration-averaged fingerprint and $\bm{\mu}$ is the dataset mean. The subscript $\bar{q}$ distinguishes the configuration-level weight from the atomic-index $j$ used in Eq.~(\ref{eq:force_peratom}).
Here $N$ is the total number of fingerprints in the dataset (individual atoms for per-atom mode, configurations for per-config mode).
The stress tensor follows analogously:
$\sigma_{\alpha\beta} = \sigma_{\alpha\beta}^{\mathrm{base}} - (k/V)\,\partial\log\det\bm{\Sigma}/\partial\varepsilon_{\alpha\beta}$.

\subsubsection{Role of the entropy weight}

The parameter~$k$ has units of energy (eV when $S_{\mathrm{local}}$ is dimensionless) and interpolates between physical sampling ($k = 0$, thermally equilibrated structures from the base model) and diversity-driven sampling ($k \gg k_BT$, highly distorted structures).
Using a grid of $k$ values produces training data spanning from near-equilibrium to highly distorted configurations.
We observed that the final learning curves are not strongly sensitive to the precise grid spacing within the range $k \in [2, 20]$~eV at $T = 1000$ to $5000$~K; the values used for the production runs are reported in Section~\ref{sec:computational}.

\subsection{Combined local + global structure generation}
\label{sec:generation}

The combined method separates exploration from selection: local entropy drives MD trajectories toward diverse configurations, while global entropy filtering ensures dataset-level information gain.

\subsubsection{Exploration phase}

For each point on a condition grid (temperatures~$T$, entropy weights~$k$, pressures~$P$, seed phases), we perform Langevin NVT MD~\cite{bussi2007canonical} (300 steps, 1~fs timestep, friction coefficient 0.01~fs$^{-1}$) using the local entropy-modified potential $E_{\mathrm{explore}}$ [Eq.~(\ref{eq:explore_energy})], driven by the atomic simulation environment~\cite{larsen2017atomic} as the MD engine.
Snapshots are collected every 10 steps, producing a pool of candidate configurations.

\subsubsection{Selection phase}

Each candidate is evaluated against the global dataset by computing the marginal information gain:
\begin{equation}
\Delta H = \log\det\bm{\Sigma}(\mathcal{D} \cup \{\mathrm{candidate}\}) - \log\det\bm{\Sigma}(\mathcal{D}) > \theta_{\min}.
\label{eq:acceptance}
\end{equation}
Only candidates exceeding the minimum gain threshold $\theta_{\min}$ are accepted.
The dataset covariance is updated incrementally upon acceptance.

\subsubsection{Refinement phase}

Near-threshold candidates (with $\theta_{\min}/2 < \Delta H < \theta_{\min}$) undergo short limited-memory BFGS optimization~\cite{liu1989limited} on the global entropy surface to maximize their information content.
If the optimized structure exceeds the threshold, it is accepted; otherwise it is discarded.
This recovers additional informative structures that were close to the acceptance boundary.

\subsection{Active learning loop}
\label{sec:active_learning}

The full pipeline is summarized in Algorithm~1.

\noindent\rule{\columnwidth}{0.6pt}\\
\noindent\textbf{Algorithm 1.}\,Combined local+global entropy-driven active learning.
Exploration uses local entropy $S_{\mathrm{local}}$ [Eq.~(\ref{eq:local_entropy})] for MD biasing; selection uses global dataset entropy gain $\Delta H$ [Eq.~(\ref{eq:acceptance})] as a post-hoc filter; refinement optimizes near-threshold candidates on the global entropy surface [Eq.~(\ref{eq:refine_energy})].\\
\noindent\rule{\columnwidth}{0.4pt}
\begin{algorithmic}[1]
\State Initialize Allegro-OAM-L foundation model as $\mathcal{M}_0$
\State Initialize fingerprint datasets $\mathcal{D}_{\mathrm{atom}},\,\mathcal{D}_{\mathrm{config}}$
\For{iteration $n = 0, 1, 2, \ldots$}
  \For{each condition $(T, k, P, \text{phase})$}
    \State Select mode: per-atom or per-config (60/40 split)
    \State \textbf{Explore}: Langevin MD with $E_{\mathrm{explore}} = E_{\mathcal{M}_n} - k \cdot S_{\mathrm{local}}$
    \State Collect candidate snapshots every 10 steps
    \For{each candidate $s$}
      \State \textbf{Select}: Compute $\Delta H(s)$ against global dataset $\mathcal{D}$
      \If{$\Delta H(s) > \theta_{\min}$}
        \State Accept $s$, update $\mathcal{D}$
      \ElsIf{$\Delta H(s) > \theta_{\min}/2$}
        \State \textbf{Refine}: LBFGS on global entropy surface
        \If{$\Delta H(s') > \theta_{\min}$}
          \State Accept $s'$, update $\mathcal{D}$
        \EndIf
      \EndIf
    \EndFor
  \EndFor
  \State Label accepted structures with DFT (energy, forces, stress)
  \State Merge with previous training data
  \State Fine-tune $\mathcal{M}_n \to \mathcal{M}_{n+1}$
  \If{validation MAE $<$ convergence threshold}
    \State \textbf{break}
  \EndIf
\EndFor
\end{algorithmic}
\noindent\rule{\columnwidth}{0.6pt}

\section{Computational details}
\label{sec:computational}

\subsection{DFT calculations}

Single-point DFT calculations are performed with VASP~\cite{kresse1996efficient,kresse1996vasp} using the projector augmented wave (PAW) method of Bl\"{o}chl~\cite{blochl1994projector} as implemented for plane-wave codes by Kresse and Joubert~\cite{kresse1999paw}.
For carbon, we use the r2SCAN~\cite{furness2020accurate} meta-GGA functional, the numerically efficient revision of the parent SCAN meta-GGA of Sun, Ruzsinszky, and Perdew~\cite{sun2015scan}, combined with the rVV10 nonlocal van der Waals correction of Sabatini, Gorni, and de Gironcoli~\cite{sabatini2013rvv10}, itself a simpler revision of the original VV10 functional of Vydrov and Van Voorhis~\cite{vydrov2010nonlocal}. The carbon-system parameter BPARAM${}=11.95$ recommended by Ning et al.~\cite{ning2022workhorse} extends the SCAN+rVV10 prescription of Peng et al.~\cite{peng2016rvv10} to the r2SCAN parent functional and correctly describes the interlayer binding in graphite that standard GGA functionals miss.
For silicon and NaCl, we use the PBE functional~\cite{perdew1996generalized}, which is adequate for these systems where van der Waals interactions play a minor role.
Common settings include a plane-wave cutoff of 600~eV, Gaussian smearing with $\sigma = 0.05$~eV, electronic convergence criterion of $10^{-6}$~eV, and $\Gamma$-centered $k$-point meshes with density 40~\AA$^{-1}$.
All calculations are single-point (NSW${}=0$) with stress computation enabled (ISIF${}=2$).

\subsubsection{Imposing pressure during NVT generation}

Because the candidate-generation MD is run in the $NVT$ ensemble (Section~\ref{sec:generation}), target pressures are imposed indirectly by pre-scaling the cell volume of each seed structure.
For each phase and target pressure $P$ we use the linear estimate $V(P) = V_0\,(1 - P/B_0)$, where $V_0$ and $B_0$ are the equilibrium volume and bulk modulus from the base MLIP, before initiating the trajectory.
The realized pressure of every accepted snapshot is computed self-consistently from its single-point DFT stress tensor and used in all downstream analysis; we therefore distinguish between the nominal (target) pressure grid that drives the volume scaling and the realized DFT pressure of the labelled structure.
The upper end of each system's nominal grid (e.g., 100~GPa for carbon) is wider than the experimental range of interest because the entropy-driven trajectory typically samples states bracketing the target, and the linear $V(P)$ estimate underestimates volume changes at the upper extreme.

\subsection{Foundation model and fine-tuning}

We use the Allegro-OAM-L foundation potential of Ref.~\onlinecite{fu2025allegro} as the base calculator.
Allegro-OAM-L is a universal MLIP based on the equivariant Allegro architecture~\cite{musaelian2023learning}, an extension of NequIP~\cite{batzner2022nequip} that replaces global message passing with strictly local equivariant tensor products to enable large-scale parallel inference; the model was pretrained on the OMat24 dataset~\cite{barroso2024omat24} and fine-tuned on MPtrj+sAlex, providing physically reasonable forces across the periodic table.

Fine-tuning updates all model parameters (no frozen layers) using the NequIP training framework~\cite{batzner2022nequip} with PyTorch Lightning, employing the Adam optimizer~\cite{kingma2014adam} (learning rate $5 \times 10^{-5}$), ReduceLROnPlateau scheduler (patience 15, factor 0.5), Huber loss~\cite{huber1964robust} with energy (weight 1.0), forces (weight 1.0), and stress (weight 0.01) coefficients, exponential moving average (decay 0.999), batch size 1 with gradient accumulation over 4 steps (effective batch size 4), gradient clipping at maximum norm 0.015 (chosen from preliminary stability tests on the foundation model where larger clip values caused training divergence), and early stopping with patience 30 epochs.
Training data are stratified by energy into bins and reweighted: low-energy near-equilibrium structures ($E < -7$~eV/atom) receive weight~3.0 to emphasize physically important configurations, intermediate structures ($-7$ to $-4$~eV/atom) weight~1.0, and high-energy distorted structures ($> -4$~eV/atom) weight~0.5.

\subsection{Fingerprint parameters}

GOM fingerprints are computed with $s$-orbital only.
The $s$-orbital GOM eigenvalue spectrum is a compact rotationally and permutationally invariant descriptor that encodes coordination number, bond lengths, and the angular distribution of neighbors via the matrix elements~\cite{zhu2016fingerprint}; we do not claim that the spectrum is uniquely invertible to atomic positions.
For local entropy during MD, we use cutoff $r_c = 3.0$~\AA, capturing the first coordination shell for carbon ($d_{\mathrm{C-C}} = 1.54$~\AA\ in diamond, 1.42~\AA\ in graphite).
For global dataset entropy, we use cutoff $r_c = 4.5$~\AA\ and dimension $d = 200$ (natx parameter), providing a richer structural descriptor that captures the second coordination shell.

The regularization parameter $\lambda = 10^{-3}$ prevents singularity of the covariance matrix when $N < d$ and contributes a constant offset to $\log\det\bm{\Sigma}$; it does not affect the marginal information gain $\Delta H$ [Eq.~(\ref{eq:acceptance})] once $N \gg d$ because $\lambda\bm{I}$ becomes negligible relative to the data covariance.
For very small datasets ($N < d$), the empirical covariance is rank-deficient and the regularization term dominates the determinant. In this regime, any candidate that adds a previously unrepresented direction in fingerprint space produces a $\Delta H$ that comfortably exceeds $\theta_{\min}$, so the global filter is effectively permissive and the early dataset accumulates rapidly until the empirical covariance becomes well-conditioned (roughly $N \gtrsim d$); a small dedicated seed pool is therefore not required to bootstrap the algorithm.
The acceptance threshold $\theta_{\min} = 0.001$ corresponds to a fractional increase of $\sim$0.1\% in the generalized variance $\det\bm{\Sigma}$ of the fingerprint distribution; structures below this threshold contribute negligibly to the dataset's information content.
In practice, $\theta_{\min}$ controls the density of accepted configurations in fingerprint space and is not sensitive to the system: the same value was used for all three systems (C, Si, NaCl) without adjustment.
Fingerprint eigenvalues are used in their raw (unnormalized) form: no standardization or whitening is applied before computing the covariance, since the absolute eigenvalue scale carries physical meaning (larger eigenvalues indicate stronger overlap, i.e., closer neighbors).
This choice could in principle bias $\log\det\bm{\Sigma}$ toward dimensions of largest absolute scale, especially for multi-component systems where species-pair eigenvalue blocks may have different numerical magnitudes; for the systems studied here this did not produce visible artifacts in the energy/pressure coverage of the accepted dataset (Section~\ref{sec:results}), but a sensitivity analysis with standardized eigenvalues and per-block rescaling is left to future work.

\subsection{Generation parameters}
\label{sec:generation_params}

We demonstrate the active learning pipeline on carbon at 0 to 100~GPa using three iterations with progressively refined generation strategies.
The same pipeline is applied independently to silicon (0 to 20~GPa, PBE, 2 seed phases: diamond cubic and $\beta$-tin) and NaCl (0 to 40~GPa, PBE, 2 seed phases: B1 rocksalt and B2 CsCl-type) for the learning curve comparisons in Section~\ref{sec:random_comparison}.

\textbf{Iteration~1} (local entropy only): 2 seed phases (diamond, graphite), $2\times 2\times 2$ supercells (16 atoms), temperatures [300, 1000, 3000]~K, $k \in \{5\}$, pressures [0, 2, 5, 10, 15, 20]~GPa, $r_c = 3.0$~\AA.
This produced 482 candidate structures, of which 238 were labeled with DFT.

\textbf{Iteration~2} (global entropy only): 12 seed phases from the Materials Project~\cite{jain2013materials} (diamond, graphite, lonsdaleite, M-carbon, and 8 additional metastable allotropes), $2\times 2\times 2$ supercells (16 to 64 atoms), temperatures [300, 1000, 3000]~K, $k \in \{0, 2, 5, 10\}$, pressures [0, 2, 5, 10, 15, 20]~GPa, $r_c = 4.5$~\AA, $d = 200$, regularization $\lambda = 10^{-3}$, minimum gain $\theta_{\min} = 0.001$.
Mode split: 60\% per-atom, 40\% per-config.
This produced 745 candidate structures, of which 734 were labeled.

\textbf{Iteration~3} (combined local + global): Same 12 seed phases, supercells capped at 32 atoms, temperatures [300, 1000, 3000, 5000]~K, $k \in \{2, 5, 10, 20\}$, pressures [0, 5, 10, 20, 40, 60, 80, 100]~GPa, local $r_c = 3.0$~\AA, global $r_c = 4.5$~\AA\ with $d = 200$, $\theta_{\min} = 0.001$, near-threshold factor 0.5.
The condition grid spans $4 \times 4 \times 8 \times 12 = 1536$ runs.
This produced 2104 candidate structures, of which 1934 converged in DFT.

The total training dataset after three iterations comprises 2906 DFT-labeled structures (238 + 734 + 1934), split 80/10/10 into train/validation/test sets.

\section{Results}
\label{sec:results}

\subsection{Training data}

For carbon, we accumulated an entropy-driven candidate pool of 814 structures via three iterations of the active-learning loop (Algorithm 1), seeded with 12 phases from the Materials Project~\cite{jain2013materials} (cubic diamond, graphite, hexagonal diamond (lonsdaleite~\cite{frondel1967lonsdaleite}), the superhard monoclinic polymorph M-carbon~\cite{li2009mcarbon}, and 8 additional metastable allotropes), and a random-MD comparison pool of 1536 structures generated from the same condition grid with no entropy bias; both pools were labeled with r2SCAN+rVV10.
For silicon, the corresponding pools contain 645 entropy-driven and 594 random-MD structures, both labeled with PBE.
For NaCl, the pools contain 524 entropy-driven and 557 random-MD structures, also labeled with PBE.
Within the training set, structures with excessive neighbor counts (total edges ${>}\,8500$ at $r_{\max} = 7$~\AA, which would otherwise exceed our GPU memory budget) are filtered out; this primarily removes high-pressure 64-atom carbon cells from the early iterations.
Sample reweighting by energy (weight 3.0 for $E < -7$~eV/atom, 1.0 for $-7$ to $-4$~eV/atom, 0.5 for $> -4$~eV/atom) emphasizes physically relevant configurations while maintaining broad coverage.

\subsection{Parity on a representative entropy-trained model}

\begin{figure}[h]
\includegraphics[width=\columnwidth]{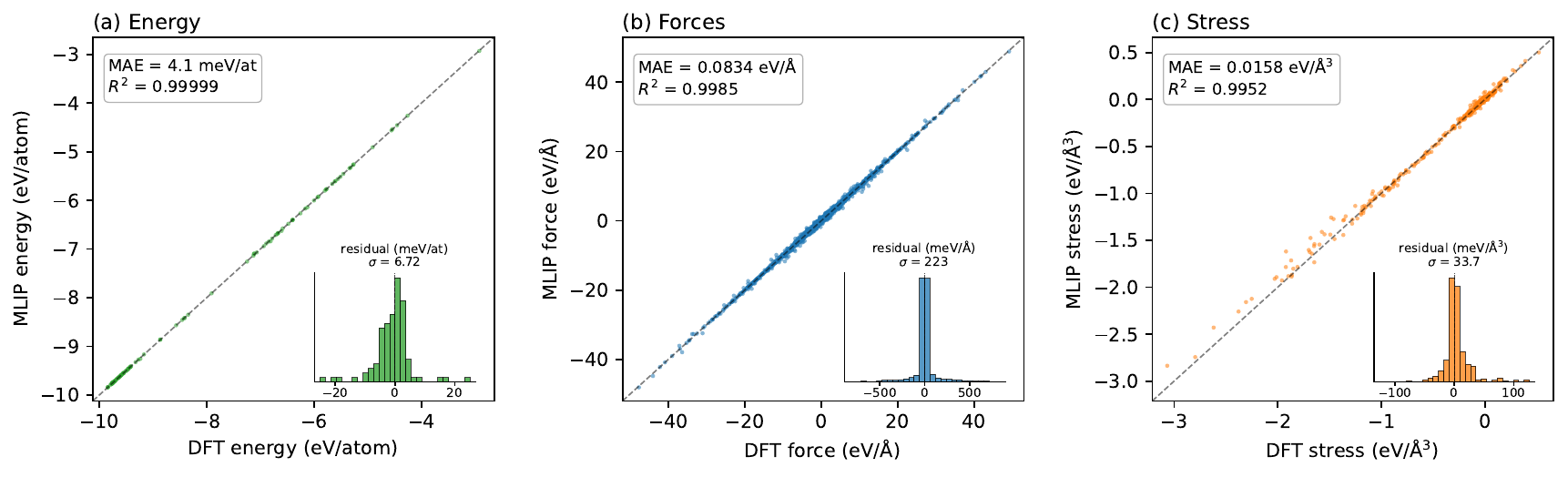}
\caption{Parity plots comparing MLIP predictions with r2SCAN+rVV10 DFT reference values on the held-out carbon test set, for a representative seed-42, $N = 800$ entropy-trained model: (a)~per-atom energy, (b)~force components, (c)~stress components. Per-panel boxes show the MAE and the coefficient of determination $R^2$.}
\label{fig:parity}
\end{figure}

Figure~\ref{fig:parity} shows parity plots comparing predictions of one representative Allegro-OAM-L model fine-tuned on $N = 800$ entropy-driven r2SCAN+rVV10 carbon structures (seed 42 of the multi-seed runs of Section~\ref{sec:random_comparison}) against DFT reference values on the held-out carbon test set.
Both the energy and force predictions track the DFT reference across the full range from near-equilibrium ($\sim -10$~eV/atom) to highly distorted ($\sim -4$~eV/atom) configurations, with stress predictions showing similar fidelity across the diagonal and off-diagonal Voigt components.
The single-seed MAE values reported on the panels (E~$=$~4.1~meV/atom, F~$=$~0.083~eV/\AA, S~$=$~0.0158~eV/\AA$^3$~$=$~2.53~GPa) lie within one standard deviation of the three-seed means reported in Table~\ref{tab:lc_carbon} at $N = 800$.

\subsection{Phonon spectra}
\label{sec:phonons}

To further validate the fine-tuned potential, we compute phonon dispersions for diamond and graphite at zero pressure using the finite displacement method as implemented in \textsc{phonopy}~\cite{togo2015first,togo2023phonopy}, with 3$\times$3$\times$3 (54 atoms) and 3$\times$3$\times$2 (72 atoms) supercells for diamond and graphite, respectively.
For each phase, the unit cell is first relaxed at zero pressure using the respective method (MLIP or DFT), and phonon frequencies are computed from force constants obtained via finite displacements of 0.01~\AA.
DFT reference phonons are computed with r2SCAN+rVV10 using VASP with tight convergence (\texttt{EDIFF}~$= 10^{-8}$~eV) and dense $k$-meshes (4$\times$4$\times$4 for diamond, 4$\times$4$\times$2 for graphite supercells).

\begin{figure}[h]
\includegraphics[width=\columnwidth]{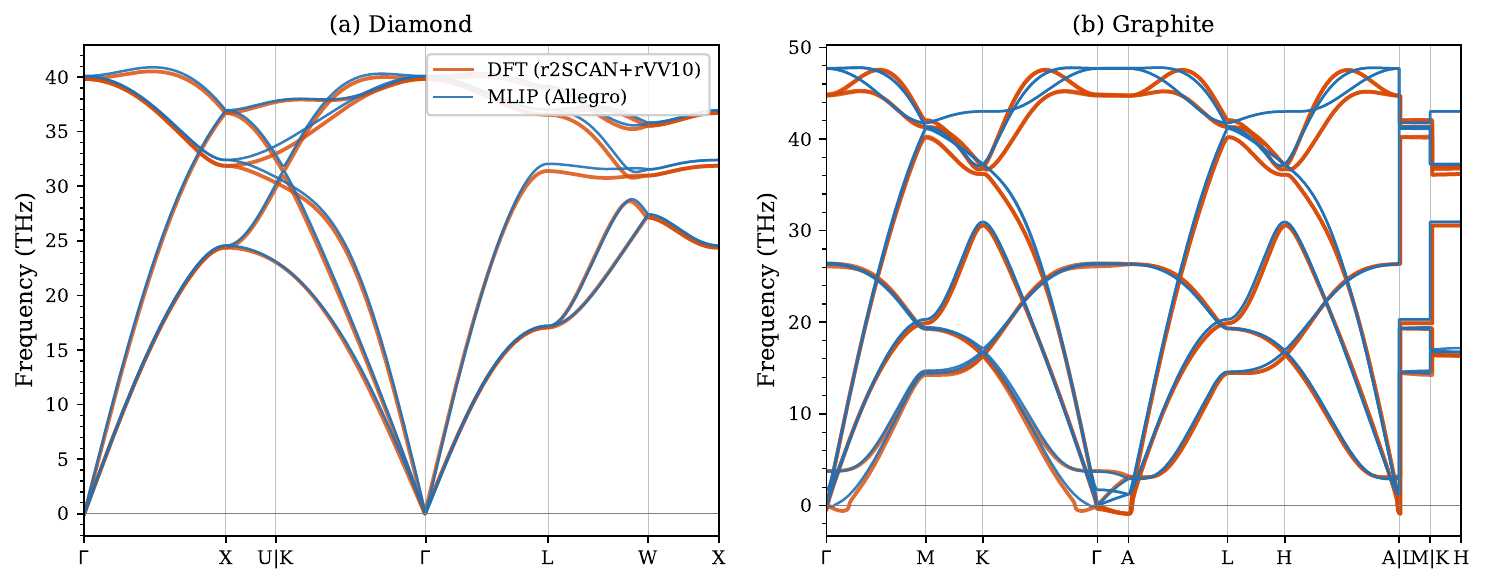}
\caption{Phonon dispersions of (a)~diamond and (b)~graphite at zero pressure. Blue lines: Allegro MLIP fine-tuned on the 814-structure entropy-driven r2SCAN+rVV10 carbon pool; orange lines: DFT r2SCAN+rVV10 reference. High-symmetry paths follow the standard conventions for FCC and hexagonal lattices.}
\label{fig:phonons}
\end{figure}

Figure~\ref{fig:phonons} shows the phonon dispersions from the MLIP overlaid with DFT reference calculations.
For diamond, the MLIP reproduces the DFT band structure with excellent accuracy: the Raman-active optical mode at $\Gamma$ is 40.9~THz (DFT: 40.5~THz; experiment: 39.9~THz~\cite{warren1967dispersion,solin1970raman}), and all acoustic and optical branches are well captured throughout the Brillouin zone.
For graphite, the MLIP correctly captures both the high-frequency in-plane modes (maximum 47.8~THz vs DFT 47.6~THz) and the low-frequency interlayer shear and breathing modes, which are sensitive to the van der Waals interaction whose magnitude in graphite has been benchmarked at the random-phase approximation level~\cite{lebegue2010cohesive}.
Small residual noise near $\Gamma$ in the graphite acoustic branches ($\sim -0.1$~THz) is negligible and reflects the flat energy landscape for interlayer sliding.
We note that the MLIP-relaxed graphite $c$-parameter (6.79~\AA) is $\sim$3\% larger than the DFT value (6.59~\AA), reflecting a slight underestimate of interlayer binding inherited from the foundation model; despite this, the phonon frequencies remain in close agreement.

\subsection{Comparison with random sampling}
\label{sec:random_comparison}

To quantify the data efficiency gained by entropy-driven sampling, we compare against a random baseline across three chemically distinct systems: carbon (r2SCAN+rVV10), silicon (PBE), and NaCl (PBE).
In each case, the random baseline uses standard NVT Langevin MD with pure MLIP forces ($k = 0$, no entropy bias), exploring the same condition grid of temperatures, pre-scaled volumes (target pressures), and seed phases as the entropy-driven pipeline.
We train separate models at four training set sizes ($N = 100, 250, 500, 800$) for both methods, using identical training hyperparameters and evaluating on a common held-out test set for each system.

\paragraph{Test set construction.}
We evaluate every $(N, \text{method}, \text{seed})$ combination on two test sets per system. The first is the in-distribution test set: the 10\% held-out partition of the 80/10/10 stratified split of the entropy-driven DFT-labeled pool (105 structures for carbon, 85 for silicon, 70 for NaCl). Within each system, both the entropy-driven and random-MD models are evaluated on the same in-distribution partition, ensuring a controlled comparison of selection strategy at a fixed test distribution. Because this partition is drawn from the entropy-driven pool, it over-represents the broad, diverse distribution that entropy-driven sampling targets and is therefore most informative about how well each model fits the distribution it was trained against.
The second is a curated independent test set (``indtest''), constructed from a separate pool that emphasizes the canonical equilibrium-and-thermal regime in which the trained MLIP will be used downstream. For each system, for each (phase $\times$ pressure), we relax the seed cell with the foundation Allegro-OAM-L (cell + atoms, $f_\text{max}=0.02$~eV/\AA) to obtain one ``equilibrium'' sample, then run two short Langevin-NVT MD trajectories (2~ps at 1000~K and 3000~K) at the relaxed cell and save five evenly spaced snapshots from the final 0.4~ps of each. All resulting cells are then DFT-labeled with the same functional and settings used for training. The phase set covers diamond, graphite, lonsdaleite, M-carbon, and a trigonal carbon polymorph (cR-32) for carbon; dc, $\beta$-tin, Imma, and sh for silicon; and B1 and B2 for NaCl. The pressure grid spans 0 to 100~GPa for carbon, 0 to 20~GPa for silicon, and 25 to 35~GPa for NaCl. The resulting indtest set sizes are 275, 219, and 86 cells for carbon, silicon, and NaCl, respectively. The indtest pool is identical for the entropy-driven and random-MD models within each system, and is independent of the train/val/test split of either training pool.
The training sets at $N = 100, 250, 500, 800$ are not nested: each is an independent random draw from its respective pool. To control for training-set draw variance and optimizer initialization variance, every $(N, \text{method})$ combination is repeated for three independent seeds (42, 123, 456); each seed governs both the random subset draw from the candidate pool and the dataloader shuffle. All numbers and error bars reported below are means $\pm$ one standard deviation over these three seeds, evaluated on a fixed global validation/test partition.

\subsubsection{Carbon}

Carbon represents a challenging covalent system where the sp$^2$/sp$^3$ bonding energetics and graphite interlayer (van der Waals) binding must be captured simultaneously.
Entropy-driven structures were drawn from the active learning pipeline described in Section~\ref{sec:active_learning}; 2112 random MD snapshots were generated, of which 1536 (73\%) converged in r2SCAN+rVV10 DFT, a lower convergence rate than the entropy-driven data (92\%), reflecting the higher fraction of strained, near-spinodal configurations that uniform random sampling produces.

Table~\ref{tab:lc_carbon} and Fig.~\ref{fig:learning_curves}(a-c) show the results.
Entropy-driven sampling outperforms random sampling at every training set size: at $N = 100$ the energy MAE is already $4.6\times$ lower (12.5 vs.\ 58.2~meV/atom), and the gap widens monotonically with $N$, reaching a $10.1\times$ ratio at $N = 800$ (4.2 vs.\ 42.8~meV/atom).
The entropy-driven energy MAE decreases monotonically with $N$ (12.5 $\to$ 8.9 $\to$ 5.9 $\to$ 4.2~meV/atom), while the random-MD energy MAE saturates near 43~meV/atom by $N = 500$; this saturation is consistent across all three seeds and indicates that adding correlated equilibrium snapshots beyond a few hundred provides limited new information.
At $N = 800$, entropy-driven sampling also achieves $3.2\times$ lower force MAE (0.084 vs.\ 0.268 eV/\AA) and $1.5\times$ lower stress MAE (2.42 vs.\ 3.75~GPa); error bars (one standard deviation over three seeds) on the entropy-driven results are below 8\% in relative terms.

\begin{table}[h]
\caption{Carbon learning curves: energy, force, and stress MAE on the common r2SCAN+rVV10 test set. Each entry is the mean over three random training-set draws (seeds 42, 123, 456), with $\pm$ one standard deviation.}
\label{tab:lc_carbon}
\begin{ruledtabular}
\begin{tabular}{lcccccc}
$N_{\mathrm{train}}$ & \multicolumn{2}{c}{Energy (meV/atom)} & \multicolumn{2}{c}{Force (eV/\AA)} & \multicolumn{2}{c}{Stress (GPa)} \\
 & Entropy & Random & Entropy & Random & Entropy & Random \\
\hline
100  & 12.5 $\pm$ 0.7  & 58.2 $\pm$ 0.4  & 0.209 $\pm$ 0.003 & 0.258 $\pm$ 0.002 & 2.36 $\pm$ 0.13 & 4.76 $\pm$ 0.23 \\
250  & 8.9 $\pm$ 0.8   & 51.6 $\pm$ 7.0  & 0.171 $\pm$ 0.006 & 0.261 $\pm$ 0.004 & 2.53 $\pm$ 0.07 & 4.32 $\pm$ 0.35 \\
500  & 5.9 $\pm$ 0.5   & 43.3 $\pm$ 1.6  & 0.131 $\pm$ 0.001 & 0.269 $\pm$ 0.004 & 2.45 $\pm$ 0.08 & 3.92 $\pm$ 0.23 \\
800  & \textbf{4.2 $\pm$ 0.2}  & 42.8 $\pm$ 1.0  & \textbf{0.084 $\pm$ 0.003} & 0.268 $\pm$ 0.004 & \textbf{2.42 $\pm$ 0.09} & 3.75 $\pm$ 0.21 \\
\end{tabular}
\end{ruledtabular}
\end{table}

\subsubsection{Silicon}

Silicon at 0 to 20~GPa undergoes the diamond cubic~$\to$~$\beta$-tin transition, involving a change from four-fold to six-fold coordination.
We generated entropy-maximized and random MD snapshots using the same pipeline, with PBE DFT labeling.

Table~\ref{tab:lc_si} and Fig.~\ref{fig:learning_curves}(d-f) show the results.
Entropy-driven sampling outperforms random sampling at every training set size: the energy MAE ratio grows from $1.7\times$ at $N = 100$ to $2.9\times$ at $N = 800$ (1.32 vs.\ 3.81~meV/atom).
The entropy-driven energy MAE decreases monotonically with $N$ (2.49 $\to$ 2.09 $\to$ 1.59 $\to$ 1.32~meV/atom), while the random-MD energy MAE is essentially flat in $N$ (4.17 $\to$ 3.81~meV/atom; differences within one standard deviation).
Stress MAE shows the same pattern, with entropy-driven sampling reaching 0.33 $\pm$ 0.01~GPa at $N = 800$ versus 0.56 $\pm$ 0.01~GPa for random.
Force MAEs are comparable between methods at small $N$ but begin to separate above $N = 500$, with entropy-driven reaching 20.0 $\pm$ 0.2~meV/\AA\ at $N = 800$.
The Si data efficiency ratio is the smallest among the three systems studied, reflecting that for this comparatively well-behaved single-element covalent system random sampling is already a reasonable baseline at the few-hundred-structure scale.

\begin{table}[h]
\caption{Silicon learning curves: energy, force, and stress MAE on a common PBE test set. Each entry is the mean over three random training-set draws (seeds 42, 123, 456), with $\pm$ one standard deviation.}
\label{tab:lc_si}
\begin{ruledtabular}
\begin{tabular}{lcccccc}
$N_{\mathrm{train}}$ & \multicolumn{2}{c}{Energy (meV/atom)} & \multicolumn{2}{c}{Force (meV/\AA)} & \multicolumn{2}{c}{Stress (GPa)} \\
 & Entropy & Random & Entropy & Random & Entropy & Random \\
\hline
100  & 2.49 $\pm$ 0.15 & 4.17 $\pm$ 0.60 & 25.7 $\pm$ 0.6 & 25.8 $\pm$ 2.3 & 0.39 $\pm$ 0.04 & 0.54 $\pm$ 0.03 \\
250  & 2.09 $\pm$ 0.43 & 4.19 $\pm$ 0.70 & 23.7 $\pm$ 1.7 & 23.5 $\pm$ 0.2 & 0.38 $\pm$ 0.03 & 0.57 $\pm$ 0.02 \\
500  & 1.59 $\pm$ 0.22 & 4.03 $\pm$ 0.34 & 21.1 $\pm$ 0.6 & 22.6 $\pm$ 0.2 & 0.35 $\pm$ 0.02 & 0.57 $\pm$ 0.01 \\
800  & \textbf{1.32 $\pm$ 0.08} & 3.81 $\pm$ 0.29 & \textbf{20.0 $\pm$ 0.2} & 22.5 $\pm$ 0.2 & \textbf{0.33 $\pm$ 0.01} & 0.56 $\pm$ 0.01 \\
\end{tabular}
\end{ruledtabular}
\end{table}

\subsubsection{NaCl}

NaCl at 0 to 40~GPa is an ionic system undergoing the B1 (rocksalt)~$\to$~B2 (CsCl-type) structural transition at $\sim$30~GPa.
This tests the method on a binary compound with long-range Coulomb interactions and a first-order reconstructive phase transition.

\begin{figure*}[t]
\includegraphics[width=0.95\textwidth]{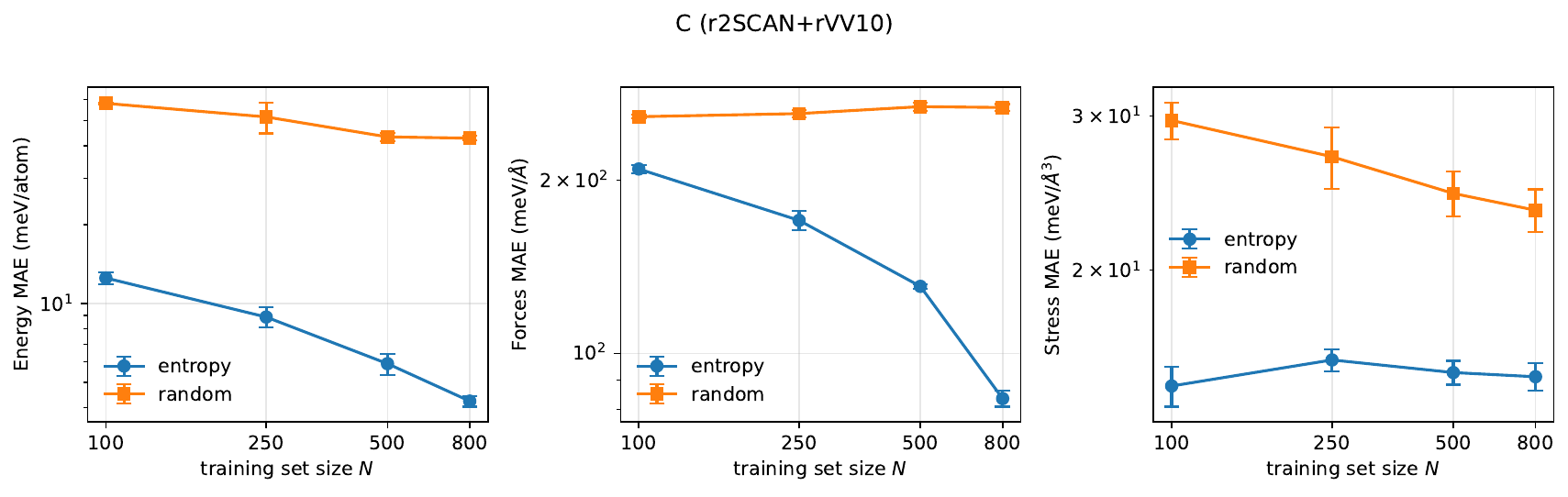}\\[2pt]
\includegraphics[width=0.95\textwidth]{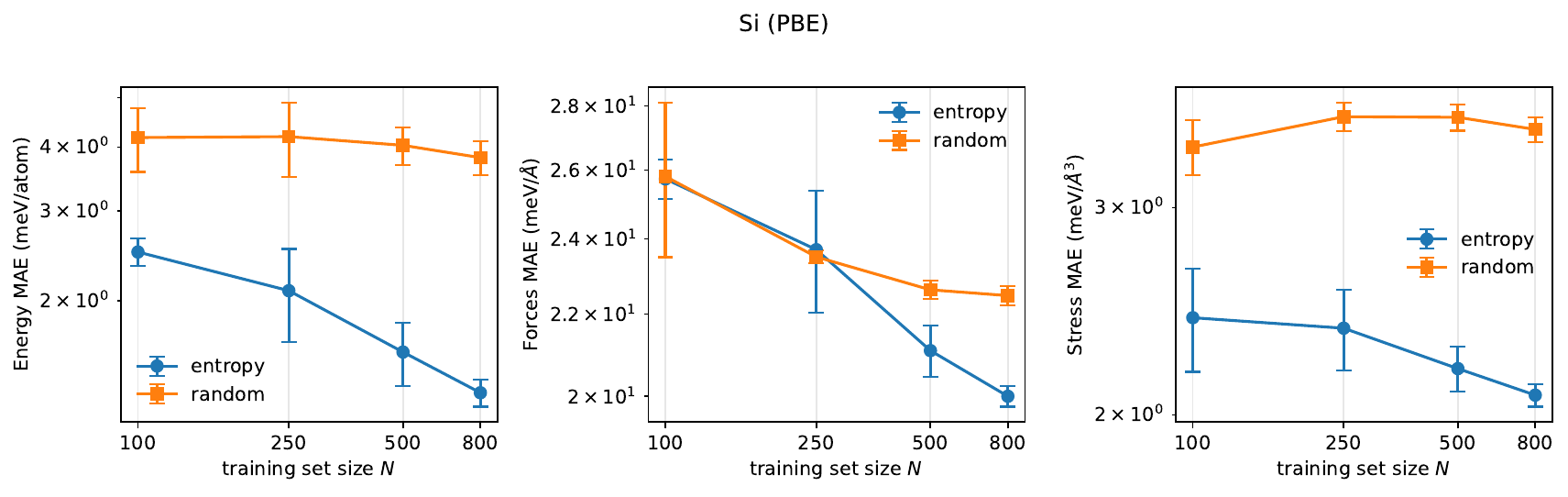}\\[2pt]
\includegraphics[width=0.95\textwidth]{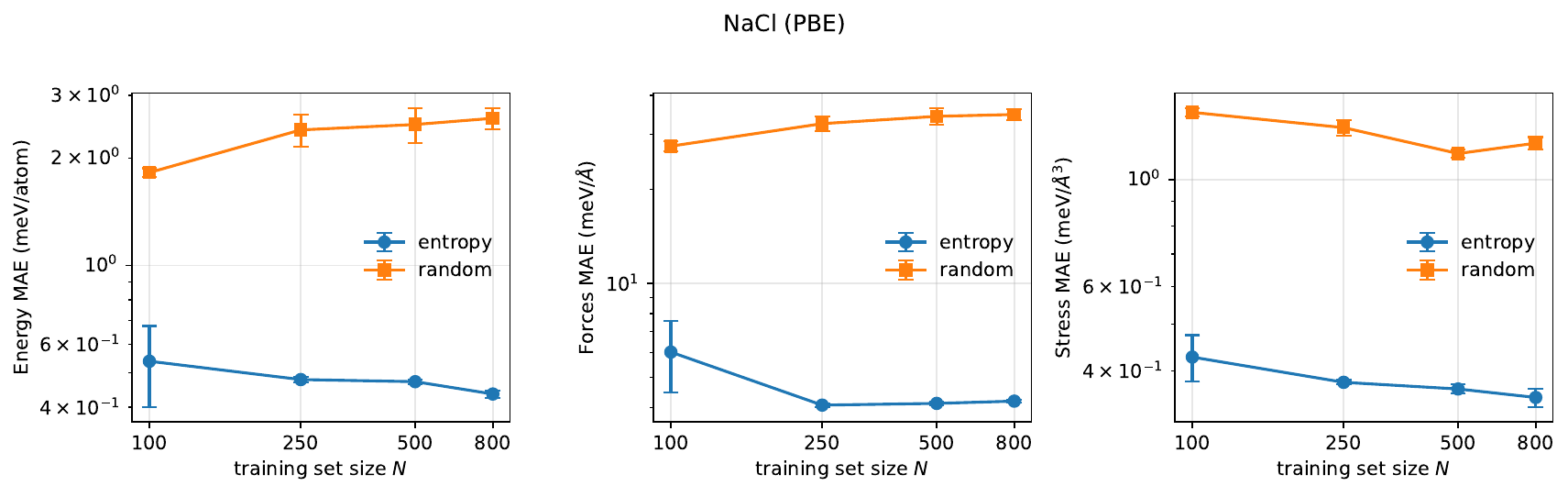}
\caption{Learning curves comparing entropy-driven (blue) and random MD (orange) sampling for (top)~carbon, (middle)~silicon, and (bottom)~NaCl on the in-distribution test sets. Each row shows, from left to right, energy MAE, force MAE, and stress MAE as a function of training set size $N \in \{100, 250, 500, 800\}$. Symbols are means over three random training-set draws (seeds 42, 123, 456); error bars are $\pm$one standard deviation. In all three systems, entropy-driven sampling outperforms random sampling above a critical training set size, while random sampling plateaus or in some cases degrades; the entropy-driven advantage exceeds one standard deviation of the random-MD result by a wide margin at $N \geq 500$.}
\label{fig:learning_curves}
\end{figure*}

Table~\ref{tab:lc_nacl} and Fig.~\ref{fig:learning_curves}(g-i) show the results.
Entropy-driven sampling reaches sub-meV/atom energy MAE already at $N = 100$ (0.54 $\pm$ 0.14~meV/atom) and remains essentially flat across the full $N = 100$ to $800$ range, with the $N = 800$ value of 0.44 $\pm$ 0.01~meV/atom only ${\sim}20\%$ better than the $N = 100$ value.
This indicates that for the comparatively simple B1/B2 NaCl configuration space, the few-hundred entropy-selected structures already span the relevant local environments with high redundancy, and additional structures yield diminishing returns.
Random sampling, in contrast, worsens with $N$ from 1.82 $\pm$ 0.05~meV/atom at $N = 100$ to 2.59 $\pm$ 0.17~meV/atom at $N = 800$, a pattern that holds across all three seeds (one-standard-deviation error bars do not overlap between $N = 100$ and $N = 800$), and so cannot be attributed to single-draw fluctuation.
We attribute this to high correlation between equilibrium snapshots from a small number of seed phases: as $N$ grows, the additional structures duplicate already-represented local environments while shifting the training distribution toward those environments, biasing the model.
At $N = 800$ the energy-MAE ratio is $5.9\times$, with comparable ratios for force ($8.3\times$) and stress ($3.2\times$).

\begin{table}[h]
\caption{NaCl learning curves: energy, force, and stress MAE on a common PBE test set. Each entry is the mean over three random training-set draws (seeds 42, 123, 456), with $\pm$ one standard deviation.}
\label{tab:lc_nacl}
\begin{ruledtabular}
\begin{tabular}{lcccccc}
$N_{\mathrm{train}}$ & \multicolumn{2}{c}{Energy (meV/atom)} & \multicolumn{2}{c}{Force (meV/\AA)} & \multicolumn{2}{c}{Stress (GPa)} \\
 & Entropy & Random & Entropy & Random & Entropy & Random \\
\hline
100  & 0.54 $\pm$ 0.14 & 1.82 $\pm$ 0.05 & 6.0 $\pm$ 1.6 & 27.5 $\pm$ 1.0 & 0.07 $\pm$ 0.01 & 0.22 $\pm$ 0.00 \\
250  & 0.48 $\pm$ 0.01 & 2.40 $\pm$ 0.25 & 4.1 $\pm$ 0.1 & 32.4 $\pm$ 1.7 & 0.06 $\pm$ 0.00 & 0.21 $\pm$ 0.01 \\
500  & 0.47 $\pm$ 0.01 & 2.48 $\pm$ 0.28 & 4.1 $\pm$ 0.1 & 34.2 $\pm$ 2.1 & 0.06 $\pm$ 0.00 & 0.18 $\pm$ 0.00 \\
800  & \textbf{0.44 $\pm$ 0.01} & 2.59 $\pm$ 0.17 & \textbf{4.2 $\pm$ 0.04} & 34.7 $\pm$ 1.4 & \textbf{0.06 $\pm$ 0.00} & 0.19 $\pm$ 0.01 \\
\end{tabular}
\end{ruledtabular}
\end{table}

\subsubsection{Generalization to a curated independent test set}
\label{sec:indtest}

\begin{figure*}[t]
\includegraphics[width=0.95\textwidth]{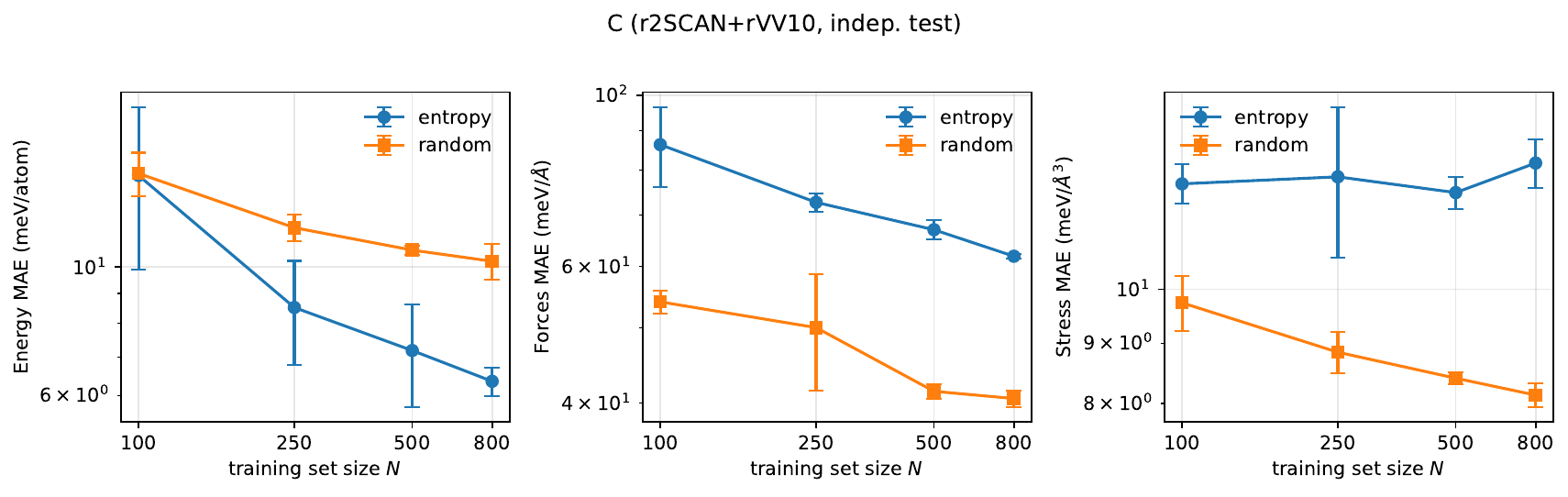}\\[2pt]
\includegraphics[width=0.95\textwidth]{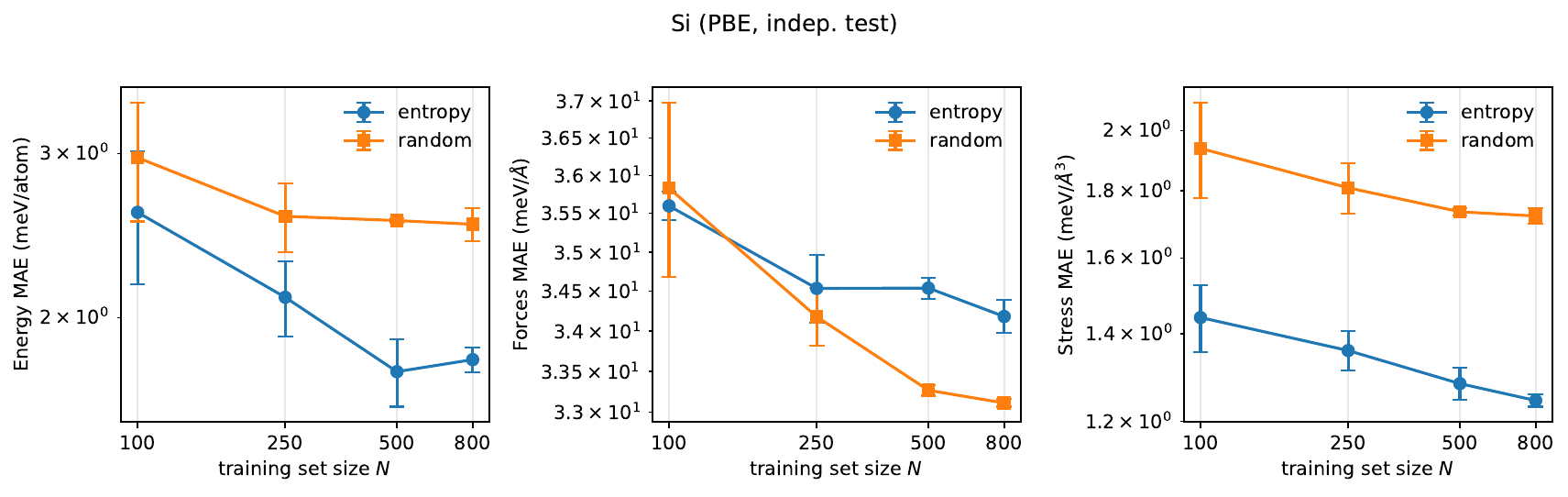}\\[2pt]
\includegraphics[width=0.95\textwidth]{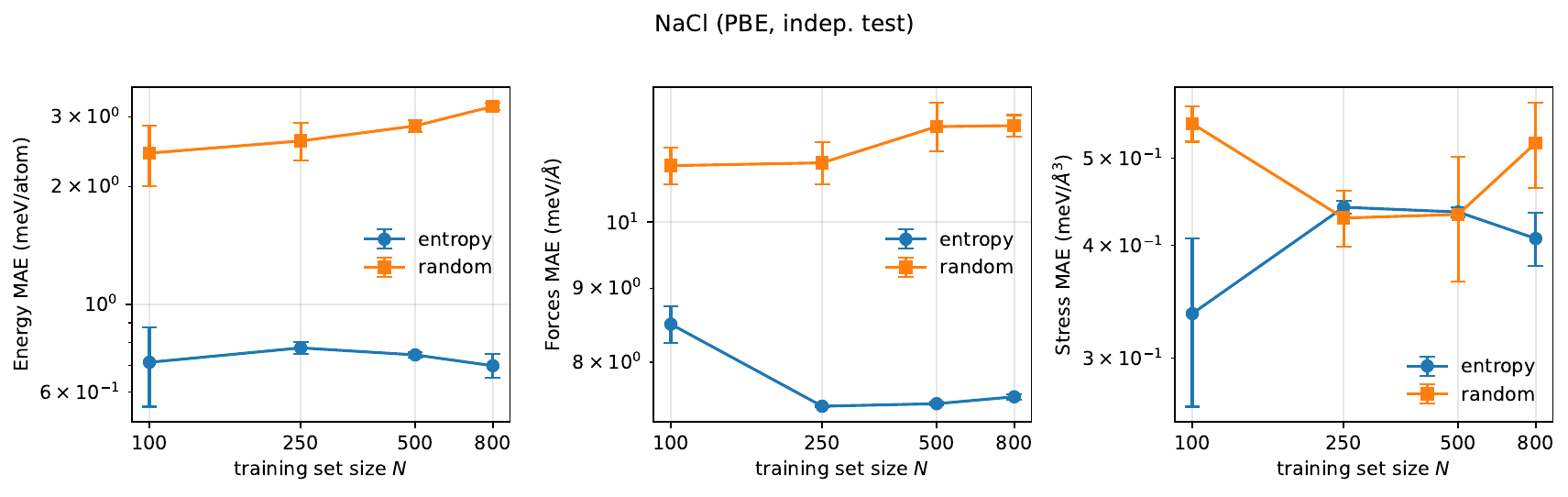}
\caption{Indtest learning curves for (top)~carbon, (middle)~silicon, and (bottom)~NaCl, in the same panel layout as Fig.~\ref{fig:learning_curves}. Symbols are means over three random training-set draws (seeds 42, 123, 456) on the curated independent test sets (275 / 219 / 86 cells). The energy panels (left column) show that entropy-driven sampling retains its advantage on the indtest for all three systems. Carbon (top row, middle and right panels) shows the reverse direction on force and stress (random-MD sampling reaches lower MAE), while silicon and NaCl follow the in-distribution trend on every channel.}
\label{fig:learning_curves_indtest}
\end{figure*}

The in-distribution learning curves above show how each method's MAE evolves on a test partition drawn from its own training pool's distribution. To probe how the resulting models generalize to the canonical equilibrium-and-thermal-MD regime that downstream MLIP use cases emphasize, we evaluated each multi-seed model on the curated indtest sets described above. Table~\ref{tab:indtest} summarizes the result at $N = 800$, and Fig.~\ref{fig:learning_curves_indtest} shows the full $N$-dependence for all three systems.

\begin{table}[h]
\caption{Indtest performance at $N = 800$. Entries are mean $\pm$ one standard deviation over three independent training-set draws (seeds 42, 123, 456) of each $(N, \text{method})$ cell. Energy in meV/atom, force in meV/\AA, stress in GPa.}
\label{tab:indtest}
\begin{ruledtabular}
\begin{tabular}{lcccccc}
System & \multicolumn{2}{c}{Energy (meV/atom)} & \multicolumn{2}{c}{Force (meV/\AA)} & \multicolumn{2}{c}{Stress (GPa)} \\
 & Entropy & Random & Entropy & Random & Entropy & Random \\
\hline
Carbon  & \textbf{6.4 $\pm$ 0.4}  & 10.2 $\pm$ 0.7  & 61.9 $\pm$ 0.5  & \textbf{40.5 $\pm$ 1.0}  & 2.05 $\pm$ 0.10 & \textbf{1.30 $\pm$ 0.03} \\
Silicon & \textbf{1.80 $\pm$ 0.06} & 2.52 $\pm$ 0.10 & 34.2 $\pm$ 0.2  & \textbf{33.1 $\pm$ 0.0}  & \textbf{0.20 $\pm$ 0.00} & 0.28 $\pm$ 0.00 \\
NaCl    & \textbf{0.70 $\pm$ 0.05} & 3.19 $\pm$ 0.07 & \textbf{7.6 $\pm$ 0.0}  & 11.7 $\pm$ 0.2  & \textbf{0.07 $\pm$ 0.00} & 0.08 $\pm$ 0.01 \\
\end{tabular}
\end{ruledtabular}
\end{table}

The energy advantage of entropy-driven sampling persists on the indtest for every system: the entropy-driven energy MAE is $1.6\times$ lower than random for carbon (6.4 vs.\ 10.2~meV/atom), $1.4\times$ lower for silicon (1.80 vs.\ 2.52~meV/atom), and $4.6\times$ lower for NaCl (0.70 vs.\ 3.19~meV/atom). Stress accuracy follows the same direction for silicon ($1.4\times$) and NaCl (within one standard deviation, with entropy slightly lower). Force accuracy is essentially tied for silicon and clearly favors entropy for NaCl ($1.5\times$).

Carbon, however, shows a qualitative reversal on force and stress: random-MD sampling reaches lower force MAE ($40.5$ vs.\ $61.9$~meV/\AA) and lower stress MAE ($1.30$ vs.\ $2.05$~GPa) at $N = 800$. The reversal is reproducible across all three seeds and is robust as $N$ grows over $100$ to $800$ (Fig.~\ref{fig:learning_curves_indtest}, top row). The most likely mechanism is that the carbon entropy pool is generated with $k$-factor up to $10$, which actively biases the trajectory toward locally diverse (hence distorted) atomic environments, while the indtest deliberately emphasizes near-equilibrium phase configurations. The entropy pool therefore under-samples the soft-vibrational regime that dominates indtest force statistics, and the random pool covers it more uniformly.
A direct test of this mechanism, and a practical remedy that recovers carbon force and stress fidelity without sacrificing the entropy-driven energy advantage, is presented in Section~\ref{sec:mixedpool}.

\subsubsection{Mixed-pool remedy for near-equilibrium force accuracy in carbon}
\label{sec:mixedpool}

The carbon force/stress reversal on the indtest (Section~\ref{sec:indtest}) is consistent with the entropy pool over-representing distorted configurations and under-representing the near-equilibrium soft-vibrational regime that the indtest emphasizes.
We test this mechanism directly by training a carbon model on a mixed pool at fixed $N = 800$: $640$ structures (80\%) drawn from the existing entropy pool plus $160$ structures (20\%) drawn from the existing random-MD pool's near-equilibrium subset, defined as the snapshots whose maximum atomic force satisfies $\max_i|\mathbf{F}_i| < 2$~eV/\AA.
Both pools are already DFT-labeled at the r2SCAN+rVV10 level used throughout this paper, so the mixed-pool experiment requires no additional electronic-structure calculations.
The validation set, test sets, model architecture, training hyperparameters, and three random training-set-draw seeds (42, 123, 456) are held identical to the multi-seed runs of Tables~\ref{tab:lc_carbon}, \ref{tab:lc_si}, and~\ref{tab:lc_nacl}.

Table~\ref{tab:mixedpool} compares the resulting mixed-pool model against the pure-entropy and pure-random pools on both the in-distribution and indtest evaluations at $N = 800$.

\begin{table}[h]
\caption{Carbon mixed-pool comparison at $N = 800$ on the in-distribution and indtest evaluations. Mixed = 80\% pure-entropy + 20\% near-equilibrium random ($\max|\mathbf{F}| < 2$~eV/\AA). Entries are mean $\pm$ one standard deviation over three independent training-set draws (seeds 42, 123, 456). Bold marks the per-channel winner within each test row.}
\label{tab:mixedpool}
\begin{ruledtabular}
\begin{tabular}{llccc}
Test set & Pool & E (meV/atom) & F (meV/\AA) & S (GPa) \\
\hline
in-dist & Entropy & \textbf{4.24 $\pm$ 0.20} & \textbf{83.5 $\pm$ 2.7}  & 2.42 $\pm$ 0.09 \\
in-dist & Random  & 42.81 $\pm$ 0.95         & 267.8 $\pm$ 3.7          & 3.75 $\pm$ 0.21 \\
in-dist & Mixed   & 9.71 $\pm$ 0.50          & 218.5 $\pm$ 1.8          & \textbf{1.77 $\pm$ 0.05} \\
indtest & Entropy & \textbf{6.4 $\pm$ 0.4}   & 61.9 $\pm$ 0.5           & 2.05 $\pm$ 0.10 \\
indtest & Random  & 10.2 $\pm$ 0.7           & \textbf{40.5 $\pm$ 1.0}  & \textbf{1.30 $\pm$ 0.03} \\
indtest & Mixed   & \textbf{6.5 $\pm$ 0.5}   & 51.0 $\pm$ 0.1           & 1.32 $\pm$ 0.02 \\
\end{tabular}
\end{ruledtabular}
\end{table}

On the indtest, the mixed-pool model matches the pure-entropy model on energy (6.5 vs.\ 6.4~meV/atom, within one standard deviation) while closing roughly half of the entropy$\to$random gap on force (51.0 vs.\ 61.9 entropy and 40.5 random) and matching the random-pool stress within noise (1.32 vs.\ 1.30~GPa, both clearly below the entropy-pool 2.05~GPa).
The variance on the mixed-pool force MAE is exceptionally low ($\pm 0.1$~meV/\AA), indicating the result is not a single-seed artifact.
On the in-distribution test, the mixed pool unsurprisingly degrades on energy and force (its training distribution is 20\% shifted away from the entropy pool whose holdout the in-distribution test draws from), but on stress the mixed pool reaches 1.77~GPa, below both the pure-entropy (2.42~GPa) and pure-random (3.75~GPa) baselines, suggesting that combining wide-distortion coverage with low-force anchors helps cell-level stress fidelity across both distributions.
The mixed-pool experiment incurred no additional DFT cost (the labels for both sub-pools were already in hand) and ${\sim}8$ hours of GPU fine-tune wallclock across the three seeds.

The reason the carbon entropy pool under-samples near-equilibrium configurations in the first place is structural rather than configurational: although the iteration-2 condition grid (Section~\ref{sec:generation_params}) included $k = 0$ (pure thermal MD) runs, the candidate snapshots from those runs are highly self-similar to structures the dataset already covers, so most of them fail the global acceptance test $\Delta H > \theta_{\min}$ and are discarded. The global log-determinant criterion intentionally discriminates against low-volume, high-probability regions of fingerprint space; the mixed-pool recipe restores the missing equilibrium mass by bypassing that filter for a small fraction of the budget.

\paragraph{Prospective recipe for new systems.}
For a new system without a pre-existing baseline random pool, we recommend the following generalization of the framework. Allocate approximately 80\% of the available DFT budget to the entropy-driven pipeline of Algorithm~1, and reserve the remaining 20\% for short Langevin NVT trajectories run with the foundation MLIP at $k = 0$ across the system's relevant temperatures and pressures, retaining only snapshots whose maximum atomic force satisfies $\max_i|\mathbf{F}_i| < 2$~eV/\AA~(or a system-appropriate near-equilibrium cutoff). These anchor snapshots should be added to the training pool without passing through the global acceptance filter, since their value is precisely that they populate the low-information equilibrium regions the global filter would otherwise discard. With this convention, the mixed pool becomes a built-in feature of the framework rather than a post-hoc correction, and the production MLIP retains both the broad-coverage advantage of entropy-driven sampling and the near-equilibrium force fidelity that downstream MD applications require.

We accordingly recommend the entropy-plus-anchor pool as the production default for downstream MD applications that demand near-equilibrium force fidelity; we revisit the implications and the open question of whether the same mixing recipe is needed for other strongly-bonded covalent systems in Section~\ref{sec:discussion}.

\subsubsection{Cross-system comparison}

Table~\ref{tab:summary} summarizes the data efficiency across all three systems.
Across the three systems, the entropy-driven energy-MAE advantage at $N = 800$ spans $2.9\times$ to $10.1\times$ on the in-distribution test, with the largest gain on carbon (where random sampling saturates at relatively high error) and the smallest on silicon (where random sampling already approximates a reasonable baseline at this scale). The energy advantage carries over to the curated indtest at smaller magnitudes ($1.4\times$ to $4.6\times$ at $N = 800$, Table~\ref{tab:indtest}), confirming that entropy-driven sampling is energy-optimal across both training and downstream-emphasized distributions for the three systems studied.
Two qualitative patterns recur on the in-distribution learning curves and are now confirmed across all three independent training-set draws: (i)~the entropy-driven energy MAE is monotonically decreasing or flat in $N$ for every system, and (ii)~the random-MD energy MAE saturates by $N = 500$ for carbon and silicon and worsens with $N$ for NaCl. The latter is most striking on NaCl, where one-standard-deviation error bars at $N = 100$ and $N = 800$ do not overlap.

\begin{table}[h]
\caption{Summary of data efficiency across three systems at $N = 800$ training structures (mean over three seeds; ratio computed from means).}
\label{tab:summary}
\begin{ruledtabular}
\begin{tabular}{lcccc}
System & Bonding & \multicolumn{2}{c}{E MAE (meV/atom)} & Ratio \\
 & type & Entropy & Random & (Random/Entropy) \\
\hline
Carbon  & covalent+vdW & 4.2   & 42.8  & 10.1$\times$ \\
Silicon & covalent     & 1.32  & 3.81  & 2.9$\times$ \\
NaCl    & ionic        & 0.44  & 2.59  & 5.9$\times$ \\
\end{tabular}
\end{ruledtabular}
\end{table}

\section{Discussion}
\label{sec:discussion}

The combined local+global entropy approach addresses both the exploration and selection challenges in training data generation.
Local entropy during MD efficiently pushes trajectories toward diverse configurations, while global entropy filtering ensures that accepted structures are non-redundant with respect to the entire accumulated dataset.
The dual per-atom/per-config modes provide a principled way to sample both ordered and disordered regions without manual intervention.

The near-threshold refinement phase recovers additional informative structures that are close to the acceptance boundary, improving data efficiency without sacrificing quality.
This is particularly effective for high-pressure structures where subtle structural differences can be physically important.

The use of a pre-trained foundation model is synergistic: Allegro-OAM-L provides physically reasonable forces that keep the MD trajectory stable even at high entropy weights ($k = 20$) and temperatures (5000~K), while the entropy term explores regions where the foundation model may be inaccurate, precisely the regions where fine-tuning data are most needed.
The iterative refinement, where each iteration's fine-tuned model serves as the base for the next, progressively improves the physical accuracy of the exploration while maintaining diversity.

The multi-seed learning curve comparisons across three chemically distinct systems (Section~\ref{sec:random_comparison}) reveal two recurring patterns.
First, the entropy-driven energy MAE decreases monotonically or remains flat with $N$ for every system tested, and the standard deviation across seeds is small enough at $N \geq 500$ ($\lesssim 10\%$ of the mean) that the trend is robust to the specific training-set draw.
Second, the random-MD energy MAE saturates by $N \approx 500$ for carbon and silicon and worsens with $N$ for NaCl, with the NaCl trend now confirmed across all three seeds (non-overlapping one-standard-deviation error bars at $N = 100$ vs $N = 800$).
We attribute the random-MD saturation/regression to redundancy among correlated equilibrium snapshots from a small set of seed phases: as $N$ grows, the additional structures duplicate already-represented local environments and bias the training distribution toward them.

\begin{figure*}[t]
\includegraphics[width=0.95\textwidth]{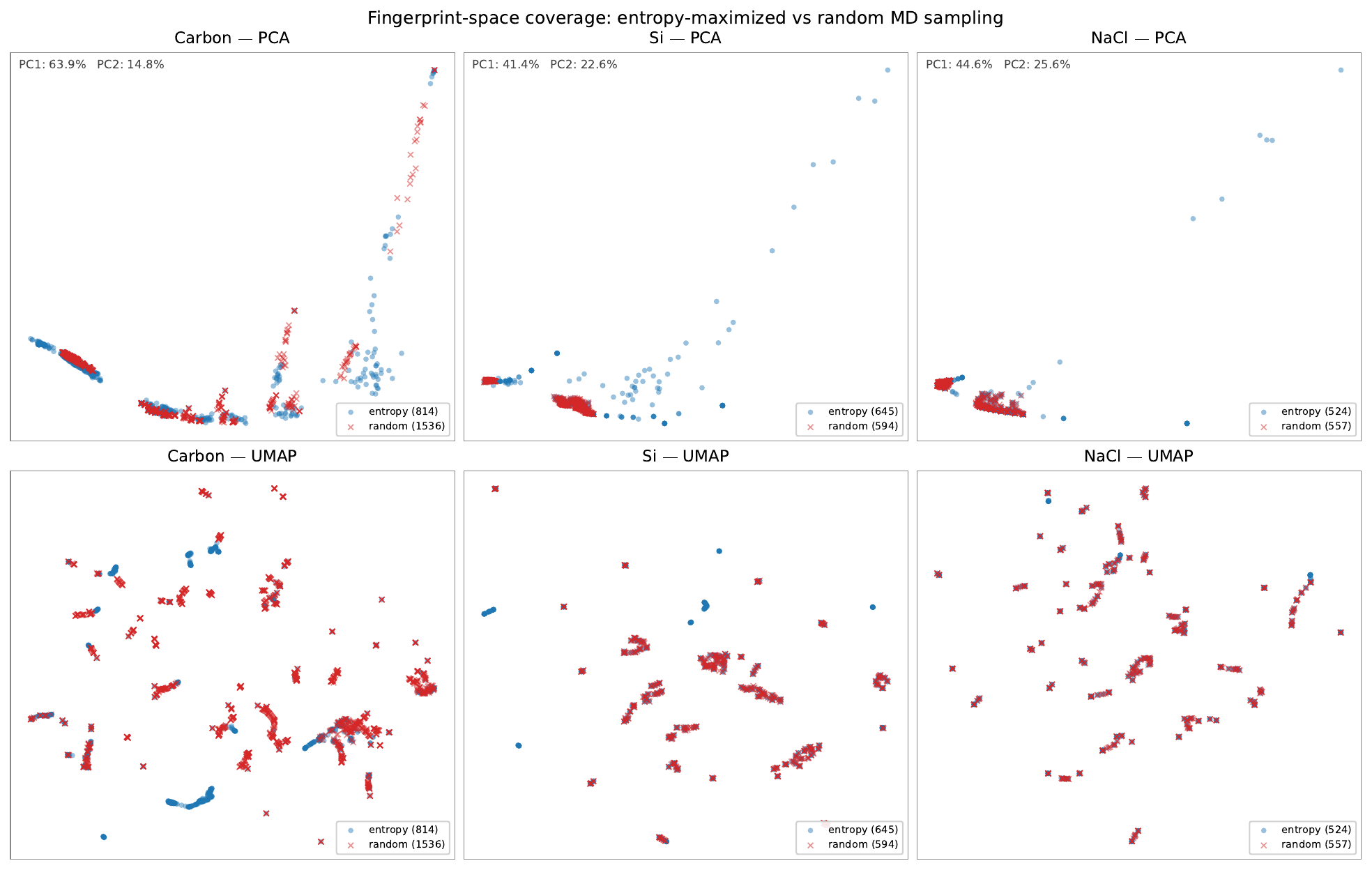}
\caption{Fingerprint-space coverage of the entropy-driven (blue) and random-MD (red) training pools for each system. Top row: PCA projections of the GOM fingerprint spectra; cumulative variance retained by PC1+PC2 is annotated on each panel. Bottom row: UMAP projections of the same fingerprints. In every panel and for every system, the entropy-driven pool occupies a substantially broader region of fingerprint space than the random-MD pool, providing a geometric account for the cross-system energy advantage in Tables~\ref{tab:lc_carbon}, \ref{tab:lc_si}, \ref{tab:lc_nacl}, and~\ref{tab:indtest}.}
\label{fig:coverage}
\end{figure*}

The data efficiency ratio at $N = 800$ depends on the system, ranging from $2.9\times$ for silicon to $10.1\times$ for carbon. The carbon ratio is largest because the random-MD baseline saturates at relatively high error ($\sim$43~meV/atom) while the entropy-driven method continues to improve, reaching 4.2~meV/atom, a regime in which subtleties of bonding (sp$^2$/sp$^3$ environments, interlayer separation) need to be resolved.
The silicon ratio is smallest because random sampling already produces a usable potential at the few-hundred-structure scale for this single-element covalent system; the entropy-driven gain is more modest in absolute terms.
NaCl falls in between, with the entropy-driven method already saturated at $N = 100$ and the random method drifting away.
The proximate mechanism for the cross-system energy advantage is visible in the configuration-space coverage of the two pools (Fig.~\ref{fig:coverage}): in both PCA and UMAP projections of the GOM fingerprint spectra, the entropy-driven pool spans a substantially broader region than the random-MD pool for all three systems, occupying directions in fingerprint space that random sampling leaves nearly empty.
This geometric breadth is what allows the entropy-trained models to interpolate well between phases and pressures with a few hundred DFT labels, while random-MD samplers, which only revisit the basins their MD trajectories already explore, accumulate redundancy without expanding coverage.

For carbon at small $N$ ($\leq 250$), entropy-driven sampling does not yet outperform random sampling on absolute energy MAE on the same scale as it does at $N = 800$, although it is consistently better at every size in the multi-seed data. The relative advantage of entropy-driven sampling therefore grows with $N$ for carbon, and a hybrid strategy that mixes a small equilibrium-MD seed pool with subsequent entropy-driven expansion is plausible for the very-low-data regime; we did not test such a hybrid here.

\paragraph{Test-distribution sensitivity and the carbon mixed-pool remedy.}
The indtest results in Section~\ref{sec:indtest} show that the energy advantage of entropy-driven sampling persists across a deliberately different test distribution for all three systems. Force and stress accuracy, however, depend on the relative weight that the test distribution places on near-equilibrium versus distorted configurations. For silicon, force accuracy is essentially tied between methods and stress accuracy still favors entropy. For NaCl, entropy retains the advantage on both. For carbon, however, the indtest reverses the in-distribution force ($1.5\times$) and stress ($1.6\times$) ranking in favor of random sampling. The mechanism is consistent with the broader fingerprint-space coverage of the entropy pool (Fig.~\ref{fig:coverage}): the carbon entropy pool was generated with $k$-factor up to $10$ and temperatures up to $5000$~K, which actively biases the trajectory toward locally diverse (hence distorted) atomic environments. The indtest emphasizes near-equilibrium phase configurations sampled with thermal MD at 1000 and 3000~K, and so the entropy pool under-samples the soft-vibrational regime that dominates indtest force statistics, while the random pool covers it more uniformly. Importantly, the integrated phase-stability quantities the paper validates, namely enthalpy ordering and phonon spectra (Section~\ref{sec:phonons}), are controlled by energy accuracy and are not affected by the carbon force/stress reversal.
The natural remedy is to combine the broad-coverage entropy pool with a small near-equilibrium anchor, and Section~\ref{sec:mixedpool} demonstrates that a $80/20$ mixed pool (entropy-driven plus low-force random-MD snapshots, $\max|\mathbf{F}|<2$~eV/\AA, at fixed total $N = 800$) carries out exactly that interpolation: at $N = 800$ on the carbon indtest, the mixed pool matches the pure-entropy energy (6.5 vs.\ 6.4~meV/atom), closes roughly half of the entropy$\to$random force gap (51.0 vs.\ 61.9 entropy and 40.5 random meV/\AA), and matches the random-pool stress (1.32 vs.\ 1.30~GPa, against entropy 2.05). The recipe is computationally free in the sense that the low-force anchor structures are already DFT-labeled in the random pool generated for the baseline; the entire experiment cost three Allegro fine-tunes and no additional electronic-structure calculations.
We therefore recommend the entropy-plus-anchor pool as a production default for downstream MD applications that demand near-equilibrium force fidelity. Whether the same mixing recipe is needed for other strongly-bonded covalent systems beyond carbon, and whether the optimal anchor fraction depends on the entropy-bias schedule used during pool generation, is left to follow-up work.

Compared to existing active learning frameworks, the present approach occupies a distinct niche.
DP-GEN~\cite{zhang2020dpgen} and related concurrent learning methods require training and querying an ensemble of models at each iteration to estimate uncertainty; this couples the selection criterion to the model architecture and adds substantial computational overhead for each exploration cycle.
Bayesian on-the-fly methods such as FLARE~\cite{vandermause2020flare} use the closed-form Gaussian-process predictive variance to flag uncertain configurations, which is elegant but constrains the surrogate to a specific (kernel-based) functional form and is less directly applicable to large foundation models.
UDD~\cite{nebgen2023udd} biases the dynamics itself toward high-uncertainty regions, providing excellent exploration but requiring differentiable uncertainty estimates integrated into the equations of motion.
D-optimal methods~\cite{podryabinkin2017active} share our log-determinant objective but formulate it in model feature space, tying the selection to the specific parametrization (e.g., moment tensor potentials).
Our descriptor-based formulation decouples the diversity criterion from the model entirely: the same GOM fingerprint covariance can be used whether the underlying potential is a linear model, a graph neural network, or a pre-trained foundation model, with no retraining or ensemble overhead during the selection phase.

\paragraph{Relation to distance-based subset selection.}
A natural alternative to the log-determinant criterion for selecting a diverse training pool from an existing set of candidate structures is a distance-based method such as farthest-point sampling (FPS) or $k$-centers clustering applied to the GOM fingerprints. These geometric criteria are simple to implement and, on a fixed candidate pool, would likely produce qualitatively similar broad-coverage selections. The log-determinant covariance criterion we adopt nevertheless has four practical advantages over distance-based selection for the active-learning setting we target. (i)~It is differentiable with respect to atomic positions through the fingerprint Jacobian, so the same objective that selects structures also provides the exploration force that drives the entropy-biased MD [Eq.~(\ref{eq:explore_energy})]; FPS-style criteria are discrete selection operators that cannot be used as MD biases. (ii)~The covariance captures the correlation structure of the dataset, not merely point-wise distances: two candidate snapshots that fall close to a densely populated direction in fingerprint space are appropriately penalized by the log-det criterion, whereas FPS would treat them as equally novel if they are far from any single existing point. (iii)~The covariance admits an incremental closed-form update via the matrix-determinant lemma, so evaluating each candidate is $O(d^2)$ regardless of dataset size; FPS-style criteria require maintaining and querying a distance structure over the accumulated dataset. (iv)~In the present pipeline the covariance is already computed for the exploration energy, so reusing it for selection is essentially free.
A controlled head-to-head comparison of log-det versus FPS or $k$-centers on a fixed pool is straightforward to set up given the GOM fingerprints we have published and is a natural target for follow-up work; we expect the gap on a fixed pool to be small but the cumulative advantage during multi-iteration active learning to be larger because of point~(i).

\paragraph{Computational overhead of the entropy-biased MD.}
The exploration phase requires, at every MD step, a GOM fingerprint evaluation per atom in the cell. For each atom this consists of a $k$-nearest-neighbor search within $r_c$ (typically $\sim$30--50 neighbors for the systems studied here) followed by diagonalization of the resulting damped overlap matrix to extract the eigenvalue spectrum; the cost scales as $O(N_{\mathrm{at}}\, n_{\mathrm{nbr}}^3)$ per step. Analytical gradients of the eigenvalues via the Hellmann-Feynman theorem avoid the expense of finite-difference estimation~\cite{zhu2016fingerprint}, and the local entropy adds an $O(N_{\mathrm{at}}^2\, d)$ pairwise nearest-neighbor distance computation in fingerprint space per step.
The incremental statistics for the global covariance are $O(d^2)$ in memory and update in $O(d^2)$ per accepted snapshot, regardless of how large the accumulated dataset grows.
For the supercell sizes used here (16--32 atoms), the per-step wall-clock cost of the entropy-biased MD was within a small constant factor of the underlying foundation-model forward pass; the exact ratio depends on hardware (the fingerprint computation runs on CPU in our implementation while the Allegro forward pass runs on GPU) and on whether the two are overlapped. A precise benchmark across hardware configurations is left to a follow-up engineering study, but we observed no regime in our experiments where the entropy-MD wall-clock became rate-limiting relative to the foundation-model evaluation itself.

\paragraph{Graphite interlayer binding.}
The fine-tuned carbon potential overpredicts the relaxed graphite $c$-axis parameter by ${\sim}3\%$ (6.79~\AA\ vs DFT 6.59~\AA), reflecting a residual underestimate of the weak, long-range vdW interlayer binding that the foundation model inherits. This is a notoriously delicate energy scale that is correctly described only by methods that include nonlocal correlation explicitly, such as the random-phase approximation~\cite{lebegue2010cohesive}.
Despite this geometric mismatch, the graphite phonon dispersion, including the interlayer shear and breathing modes that are most sensitive to the van der Waals interaction, is well reproduced (Section~\ref{sec:phonons}).
The structural mismatch is therefore tolerable for transition-pressure prediction but would limit applications that depend on exact interlayer geometries (e.g., interlayer-sliding barriers in graphite-derived stackings).
A targeted iteration with additional graphite-stacking configurations sampled at the entropy-driven step, or an explicit interlayer-binding curve included in the validation suite, would address this.

\section{Conclusions}
\label{sec:conclusions}

We have presented a combined local+global entropy-maximized active learning framework for generating training data for machine-learned interatomic potentials.
The method uses local entropy-biased MD for exploration, global covariance-based entropy for dataset-aware selection, dual-mode fingerprint tracking for ordered/disordered coverage, near-threshold refinement for improved data efficiency, and a pre-trained foundation model (Allegro-OAM-L) for physically grounded sampling.
Learning curve comparisons across three chemically distinct systems, namely carbon (covalent, r2SCAN+rVV10), silicon (covalent/metallic, PBE), and NaCl (ionic, PBE), averaged over three independent training-set draws per condition, show that entropy-driven sampling achieves a factor of $2.9$ to $10.1$ lower energy MAE than random MD sampling at $N = 800$ on the in-distribution holdout, with system-dependent magnitude.
The energy advantage carries over to a curated independent test set of equilibrium and thermal-MD configurations across each system's relevant phases and pressures ($1.4\times$ to $4.6\times$ at $N = 800$), confirming that the data-efficiency gain is not an artifact of evaluating on the same distribution as the training pool.
Two qualitative patterns recur across all systems on the in-distribution learning curves: the entropy-driven energy MAE is monotonically decreasing or flat in $N$, while the random-MD energy MAE either saturates or worsens with $N$ as additional correlated equilibrium snapshots provide diminishing or negative returns.
Force and stress accuracy on the curated independent test set is more test-distribution dependent: entropy-driven sampling matches or improves on random for silicon and NaCl, but underperforms random for carbon, consistent with the broader fingerprint-space coverage of the entropy pool (Fig.~\ref{fig:coverage}) over-representing distorted configurations relative to the near-equilibrium structures emphasized by the curated test.
For carbon, an $80/20$ entropy-plus-anchor pool (obtained at no additional DFT cost by drawing 20\% of the training cells from the existing random pool's low-force subset) matches pure entropy on indtest energy while recovering random-pool force and stress accuracy (Section~\ref{sec:mixedpool}), resolving the reversal and providing a methodological recommendation for downstream MD applications.
The fine-tuned carbon potential reproduces DFT phonon dispersions for diamond and graphite (Section~\ref{sec:phonons}), validating the physical quality of the entropy-driven training data.
The approach is general and applicable to any elemental or multi-component system accessible by DFT, including systems labeled with beyond-DFT methods such as dynamical mean-field theory~\cite{kotliar2006electronic} where each label is orders of magnitude more expensive than a DFT single point and data efficiency is correspondingly more critical.

\bmhead{Acknowledgements}
This work was supported by the startup funds of the office of the Dean of SASN of Rutgers University-Newark.
The authors acknowledge the Office of Advanced Research Computing (OARC) at Rutgers for providing access to the Amarel cluster and associated research computing resources.

\bibliography{biblio}

@article{behler2016perspective,
  author  = {Behler, J\"{o}rg},
  title   = {Perspective: Machine learning potentials for atomistic simulations},
  journal = {J. Chem. Phys.},
  volume  = {145},
  pages   = {170201},
  year    = {2016},
}

@article{deringer2019machine,
  author  = {Deringer, Volker L. and Caro, Miguel A. and Cs\'{a}nyi, G\'{a}bor},
  title   = {Machine learning interatomic potentials as emerging tools for materials science},
  journal = {Adv. Mater.},
  volume  = {31},
  pages   = {1902765},
  year    = {2019},
}

@article{unke2021machine,
  author  = {Unke, Oliver T. and Chmiela, Stefan and Sauceda, Huziel E. and Gastegger, Michael and Poltavsky, Igor and Sch\"{u}tt, Kristof T. and Tkatchenko, Alexandre and M\"{u}ller, Klaus-Robert},
  title   = {Machine learning force fields},
  journal = {Chem. Rev.},
  volume  = {121},
  pages   = {10142},
  year    = {2021},
}

@misc{yang2024mattersim,
  author        = {Yang, Han and Hu, Chenxi and Zhou, Yichi and Liu, Xixian and Shi, Yu and Li, Jielan and Li, Guanzhi and Chen, Zekun and Chen, Shuizhou and Zeni, Claudio and Horton, Matthew and Pinsler, Robert and Fowler, Andrew and Z\"{u}gner, Daniel and Xie, Tian and Smith, Jake and Sun, Lixin and Wang, Qian and Kong, Lingyu and Liu, Chang and Hao, Hongxia and Lu, Ziheng},
  title         = {{MatterSim}: A deep learning atomistic model across elements, temperatures and pressures},
  year          = {2024},
  eprint        = {2405.04967},
  archivePrefix = {arXiv},
  primaryClass  = {cond-mat.mtrl-sci},
}

@misc{batatia2024foundation,
  author        = {Batatia, Ilyes and Benner, Philipp and Chiang, Yuan and Elena, Alin M. and Kov\'{a}cs, D\'{a}vid P. and Riebesell, Janosh and Advincula, Xavier R. and Asta, Mark and Avaylon, Matthew and Baldwin, William J. and others},
  title         = {A foundation model for atomistic materials chemistry},
  year          = {2024},
  eprint        = {2401.00096},
  archivePrefix = {arXiv},
  primaryClass  = {physics.chem-ph},
}

@article{deng2023chgnet,
  author  = {Deng, Bowen and Zhong, Peichen and Jun, KyuJung and Riebesell, Janosh and Han, Kevin and Bartel, Christopher J. and Ceder, Gerbrand},
  title   = {{CHGNet} as a pretrained universal neural network potential for charge-informed atomistic modelling},
  journal = {Nat. Mach. Intell.},
  volume  = {5},
  pages   = {1031},
  year    = {2023},
}

@article{karabin2020entropy,
  author  = {Karabin, Mariia and Perez, Danny},
  title   = {An entropy-maximization approach to automated training set generation for interatomic potentials},
  journal = {J. Chem. Phys.},
  volume  = {153},
  pages   = {094110},
  year    = {2020},
}

@article{subramanyam2025,
  author  = {Subramanyam, Aparna P. A. and Perez, Danny},
  title   = {Information-maximization based active learning of interatomic potentials},
  journal = {npj Comput. Mater.},
  volume  = {11},
  pages   = {218},
  year    = {2025},
}

@article{zhu2016fingerprint,
  author  = {Zhu, Li and Amsler, Maximilian and Fuhrer, Tobias and Schaefer, Bastian and Faraji, Somayeh and Rostami, Samare and Ghasemi, S. Alireza and Sadeghi, Ali and Grauzinyte, Migle and Wolverton, Chris and Goedecker, Stefan},
  title   = {A fingerprint based metric for measuring similarities of crystalline structures},
  journal = {J. Chem. Phys.},
  volume  = {144},
  pages   = {034203},
  year    = {2016},
}

@article{kresse1996efficient,
  author  = {Kresse, G. and Furthm\"{u}ller, J.},
  title   = {Efficient iterative schemes for {\it ab initio} total-energy calculations using a plane-wave basis set},
  journal = {Phys. Rev. B},
  volume  = {54},
  pages   = {11169},
  year    = {1996},
}

@article{kresse1996vasp,
  author  = {Kresse, G. and Furthm\"{u}ller, J.},
  title   = {Efficiency of ab-initio total energy calculations for metals and semiconductors using a plane-wave basis set},
  journal = {Comput. Mater. Sci.},
  volume  = {6},
  pages   = {15},
  year    = {1996},
}

@article{blochl1994projector,
  author  = {Bl\"{o}chl, P. E.},
  title   = {Projector augmented-wave method},
  journal = {Phys. Rev. B},
  volume  = {50},
  pages   = {17953},
  year    = {1994},
}

@article{perdew1996generalized,
  author  = {Perdew, John P. and Burke, Kieron and Ernzerhof, Matthias},
  title   = {Generalized gradient approximation made simple},
  journal = {Phys. Rev. Lett.},
  volume  = {77},
  pages   = {3865},
  year    = {1996},
}

@article{furness2020accurate,
  author  = {Furness, James W. and Kaplan, Aaron D. and Ning, Jinliang and Perdew, John P. and Sun, Jianwei},
  title   = {Accurate and numerically efficient {r$^2$SCAN} meta-generalized gradient approximation},
  journal = {J. Phys. Chem. Lett.},
  volume  = {11},
  pages   = {8208},
  year    = {2020},
}

@article{ning2022workhorse,
  author  = {Ning, Jinliang and Kothakonda, Manish and Furness, James W. and Kaplan, Aaron D. and Ehlert, Sebastian and Brandenburg, Jan Gerit and Perdew, John P. and Sun, Jianwei},
  title   = {Workhorse minimally empirical dispersion-corrected density functional with tests for weakly bound systems: {r$^2$SCAN+rVV10}},
  journal = {Phys. Rev. B},
  volume  = {106},
  pages   = {075422},
  year    = {2022},
}

@misc{fu2025allegro,
  author        = {Tan, Chuin Wei and Descoteaux, Marc L. and Kotak, Mit and de Miranda Nascimento, Gabriel and Kavanagh, Se\'{a}n R. and Zichi, Laura and Wang, Menghang and Saluja, Aadit and Hu, Yizhong R. and Smidt, Tess and Johansson, Anders and Witt, William C. and Kozinsky, Boris and Musaelian, Albert},
  title         = {High-performance training and inference for deep equivariant interatomic potentials},
  year          = {2025},
  eprint        = {2504.16068},
  archivePrefix = {arXiv},
  primaryClass  = {physics.comp-ph},
  note          = {Releases the Allegro-OAM-L foundation potential; pretrained on OMat24 and fine-tuned on MPtrj+sAlex.},
}

@misc{barroso2024omat24,
  author        = {Barroso-Luque, Luis and Shuaibi, Muhammed and Fu, Xiang and Wood, Brandon M. and Dzamba, Misko and Gao, Meng and Rizvi, Ammar and Zitnick, C. Lawrence and Ulissi, Zachary W.},
  title         = {Open {M}aterials 2024 ({OMat24}) inorganic materials dataset and models},
  year          = {2024},
  eprint        = {2410.12771},
  archivePrefix = {arXiv},
  primaryClass  = {cond-mat.mtrl-sci},
}

@article{musaelian2023learning,
  author  = {Musaelian, Albert and Batzner, Simon and Johansson, Anders and Sun, Lixin and Owen, Cameron J. and Kornbluth, Mordechai and Kozinsky, Boris},
  title   = {Learning local equivariant representations for large-scale atomistic dynamics},
  journal = {Nat. Commun.},
  volume  = {14},
  pages   = {579},
  year    = {2023},
}

@article{togo2015first,
  author  = {Togo, Atsushi and Tanaka, Isao},
  title   = {First principles phonon calculations in materials science},
  journal = {Scr. Mater.},
  volume  = {108},
  pages   = {1},
  year    = {2015},
}

@article{warren1967dispersion,
  author  = {Warren, J. L. and Yarnell, J. L. and Dolling, G. and Cowley, R. A.},
  title   = {Lattice dynamics of diamond},
  journal = {Phys. Rev.},
  volume  = {158},
  pages   = {805},
  year    = {1967},
}

@article{zhang2020dpgen,
  author  = {Zhang, Yuzhi and Wang, Haidi and Chen, Weijie and Zeng, Jinzhe and Zhang, Linfeng and Wang, Han and E, Weinan},
  title   = {{DP-GEN}: A concurrent learning platform for the generation of reliable deep learning based potential energy models},
  journal = {Comput. Phys. Commun.},
  volume  = {253},
  pages   = {107206},
  year    = {2020},
}

@article{nebgen2023udd,
  author  = {Kulichenko, Maksim and Barros, Kipton and Lubbers, Nicholas and Li, Ying Wai and Messerly, Richard and Tretiak, Sergei and Smith, Justin S. and Nebgen, Benjamin},
  title   = {Uncertainty-driven dynamics for active learning of interatomic potentials},
  journal = {Nat. Comput. Sci.},
  volume  = {4},
  pages   = {29},
  year    = {2024},
}

@article{podryabinkin2017active,
  author  = {Podryabinkin, Evgeny V. and Shapeev, Alexander V.},
  title   = {Active learning of linearly parametrized interatomic potentials},
  journal = {Comput. Mater. Sci.},
  volume  = {140},
  pages   = {171},
  year    = {2017},
}

@article{shapeev2017moment,
  author  = {Shapeev, Alexander V.},
  title   = {Moment tensor potentials: A class of systematically improvable interatomic potentials},
  journal = {Multiscale Model. Simul.},
  volume  = {14},
  pages   = {1153},
  year    = {2016},
}

@article{batzner2022nequip,
  author  = {Batzner, Simon and Musaelian, Albert and Sun, Lixin and Geiger, Mario and Mailoa, Jonathan P. and Kornbluth, Mordechai and Molinari, Nicola and Smidt, Tess E. and Kozinsky, Boris},
  title   = {{E(3)}-equivariant graph neural networks for data-efficient and accurate interatomic potentials},
  journal = {Nat. Commun.},
  volume  = {13},
  pages   = {2453},
  year    = {2022},
}

@inproceedings{batatia2022mace,
  author    = {Batatia, Ilyes and Kov\'{a}cs, D\'{a}vid P. and Simm, Gregor N. C. and Ortner, Christoph and Cs\'{a}nyi, G\'{a}bor},
  title     = {{MACE}: Higher order equivariant message passing neural networks for fast and accurate force fields},
  booktitle = {Advances in Neural Information Processing Systems},
  volume    = {35},
  pages     = {11423},
  year      = {2022},
}

@article{bartok2010gap,
  author  = {Bart\'{o}k, Albert P. and Payne, Mike C. and Kondor, Risi and Cs\'{a}nyi, G\'{a}bor},
  title   = {{Gaussian} approximation potentials: The accuracy of quantum mechanics, without the electrons},
  journal = {Phys. Rev. Lett.},
  volume  = {104},
  pages   = {136403},
  year    = {2010},
}

@article{merchant2023gnome,
  author  = {Merchant, Amil and Batzner, Simon and Schoenholz, Samuel S. and Aykol, Muratahan and Cheon, Gowoon and Cubuk, Ekin Dogus},
  title   = {Scaling deep learning for materials discovery},
  journal = {Nature},
  volume  = {624},
  pages   = {80},
  year    = {2023},
}

@article{vandermause2020flare,
  author  = {Vandermause, Jonathan and Torrisi, Steven B. and Batzner, Simon and Xie, Yu and Sun, Lixin and Kolpak, Alexie M. and Kozinsky, Boris},
  title   = {On-the-fly active learning of interpretable {Bayesian} force fields for atomistic rare events},
  journal = {npj Comput. Mater.},
  volume  = {6},
  pages   = {20},
  year    = {2020},
}

@article{sabatini2013rvv10,
  author  = {Sabatini, Riccardo and Gorni, Tommaso and de Gironcoli, Stefano},
  title   = {Nonlocal van der {Waals} density functional made simple and efficient},
  journal = {Phys. Rev. B},
  volume  = {87},
  pages   = {041108(R)},
  year    = {2013},
}

@article{jain2013materials,
  author  = {Jain, Anubhav and Ong, Shyue Ping and Hautier, Geoffroy and Chen, Wei and Richards, William Davidson and Dacek, Stephen and Cholia, Shreyas and Gunter, Dan and Skinner, David and Ceder, Gerbrand and Persson, Kristin A.},
  title   = {Commentary: The {Materials Project}: A materials genome approach to accelerating materials innovation},
  journal = {APL Mater.},
  volume  = {1},
  pages   = {011002},
  year    = {2013},
}

@article{togo2023phonopy,
  author  = {Togo, Atsushi and Chaput, Laurent and Tadano, Terumasa and Tanaka, Isao},
  title   = {Implementation strategies in {phonopy} and {phono3py}},
  journal = {J. Phys. Condens. Matter},
  volume  = {35},
  pages   = {353001},
  year    = {2023},
}

@article{behler2007generalized,
  author  = {Behler, J\"{o}rg and Parrinello, Michele},
  title   = {Generalized neural-network representation of high-dimensional potential-energy surfaces},
  journal = {Phys. Rev. Lett.},
  volume  = {98},
  pages   = {146401},
  year    = {2007},
}

@article{schutt2018schnet,
  author  = {Sch\"{u}tt, K. T. and Sauceda, H. E. and Kindermans, P.-J. and Tkatchenko, A. and M\"{u}ller, K.-R.},
  title   = {{SchNet}: A deep learning architecture for molecules and materials},
  journal = {J. Chem. Phys.},
  volume  = {148},
  pages   = {241722},
  year    = {2018},
}

@article{drautz2019ace,
  author  = {Drautz, Ralf},
  title   = {Atomic cluster expansion for accurate and transferable interatomic potentials},
  journal = {Phys. Rev. B},
  volume  = {99},
  pages   = {014104},
  year    = {2019},
}

@article{bartok2013soap,
  author  = {Bart\'{o}k, Albert P. and Kondor, Risi and Cs\'{a}nyi, G\'{a}bor},
  title   = {On representing chemical environments},
  journal = {Phys. Rev. B},
  volume  = {87},
  pages   = {184115},
  year    = {2013},
}

@article{wang2018deepmd,
  author  = {Wang, Han and Zhang, Linfeng and Han, Jiequn and E, Weinan},
  title   = {{DeePMD-kit}: A deep learning package for many-body potential energy representation and molecular dynamics},
  journal = {Comput. Phys. Commun.},
  volume  = {228},
  pages   = {178},
  year    = {2018},
}

@article{vydrov2010nonlocal,
  author  = {Vydrov, Oleg A. and Van Voorhis, Troy},
  title   = {Nonlocal van der {Waals} density functional: The simpler the better},
  journal = {J. Chem. Phys.},
  volume  = {133},
  pages   = {244103},
  year    = {2010},
}

@article{peng2016rvv10,
  author  = {Peng, Haowei and Yang, Zeng-Hui and Perdew, John P. and Sun, Jianwei},
  title   = {Versatile van der {Waals} density functional based on a meta-generalized gradient approximation},
  journal = {Phys. Rev. X},
  volume  = {6},
  pages   = {041005},
  year    = {2016},
}

@article{sun2015scan,
  author  = {Sun, Jianwei and Ruzsinszky, Adrienn and Perdew, John P.},
  title   = {Strongly constrained and appropriately normed semilocal density functional},
  journal = {Phys. Rev. Lett.},
  volume  = {115},
  pages   = {036402},
  year    = {2015},
}

@article{kresse1999paw,
  author  = {Kresse, G. and Joubert, D.},
  title   = {From ultrasoft pseudopotentials to the projector augmented-wave method},
  journal = {Phys. Rev. B},
  volume  = {59},
  pages   = {1758},
  year    = {1999},
}

@article{larsen2017atomic,
  author  = {Hjorth Larsen, Ask and J{\o}rgen Mortensen, Jens and Blomqvist, Jakob and Castelli, Ivano E. and Christensen, Rune and Du\l{}ak, Marcin and Friis, Jesper and Groves, Michael N. and Hammer, Bj{\o}rk and Hargus, Cory and Hermes, Eric D. and Jennings, Paul C. and Bjerre Jensen, Peter and Kermode, James and Kitchin, John R. and Leonhard Kolsbjerg, Esben and Kubal, Joseph and Kaasbjerg, Kristen and Lysgaard, Steen and Bergmann Maronsson, J\'{o}n and Maxson, Tristan and Olsen, Thomas and Pastewka, Lars and Peterson, Andrew and Rostgaard, Carsten and Schi{\o}tz, Jakob and Sch\"{u}tt, Ole and Strange, Mikkel and Thygesen, Kristian S. and Vegge, Tejs and Vilhelmsen, Lasse and Walter, Michael and Zeng, Zhenhua and Jacobsen, Karsten W.},
  title   = {The atomic simulation environment---a {Python} library for working with atoms},
  journal = {J. Phys. Condens. Matter},
  volume  = {29},
  pages   = {273002},
  year    = {2017},
}

@article{bussi2007canonical,
  author  = {Bussi, Giovanni and Donadio, Davide and Parrinello, Michele},
  title   = {Canonical sampling through velocity rescaling},
  journal = {J. Chem. Phys.},
  volume  = {126},
  pages   = {014101},
  year    = {2007},
}

@article{liu1989limited,
  author  = {Liu, Dong C. and Nocedal, Jorge},
  title   = {On the limited memory {BFGS} method for large scale optimization},
  journal = {Math. Program.},
  volume  = {45},
  pages   = {503},
  year    = {1989},
}

@misc{kingma2014adam,
  author        = {Kingma, Diederik P. and Ba, Jimmy},
  title         = {{Adam}: A method for stochastic optimization},
  year          = {2014},
  eprint        = {1412.6980},
  archivePrefix = {arXiv},
  primaryClass  = {cs.LG},
  note          = {Published as a conference paper at ICLR 2015.},
}

@article{bernstein2019denovo,
  author  = {Bernstein, Noam and Cs\'{a}nyi, G\'{a}bor and Deringer, Volker L.},
  title   = {De novo exploration and self-guided learning of potential-energy surfaces},
  journal = {npj Comput. Mater.},
  volume  = {5},
  pages   = {99},
  year    = {2019},
}

@article{schran2020committee,
  author  = {Schran, Christoph and Brezina, Krystof and Marsalek, Ondrej},
  title   = {Committee neural network potentials control generalization errors and enable active learning},
  journal = {J. Chem. Phys.},
  volume  = {153},
  pages   = {104105},
  year    = {2020},
}

@article{li2009mcarbon,
  author  = {Li, Quan and Ma, Yanming and Oganov, Artem R. and Wang, Hongbo and Wang, Hui and Xu, Yu and Cui, Tian and Mao, Ho-Kwang and Zou, Guangtian},
  title   = {Superhard monoclinic polymorph of carbon},
  journal = {Phys. Rev. Lett.},
  volume  = {102},
  pages   = {175506},
  year    = {2009},
}

@article{lebegue2010cohesive,
  author  = {Leb\`{e}gue, S. and Harl, J. and Gould, T. and \'{A}ngy\'{a}n, J. G. and Kresse, G. and Dobson, J. F.},
  title   = {Cohesive properties and asymptotics of the dispersion interaction in graphite by the random phase approximation},
  journal = {Phys. Rev. Lett.},
  volume  = {105},
  pages   = {196401},
  year    = {2010},
}

@article{frondel1967lonsdaleite,
  author  = {Frondel, Clifford and Marvin, Ursula B.},
  title   = {Lonsdaleite, a hexagonal polymorph of diamond},
  journal = {Nature},
  volume  = {214},
  pages   = {587},
  year    = {1967},
}

@article{kotliar2006electronic,
  author  = {Kotliar, G. and Savrasov, S. Y. and Haule, K. and Oudovenko, V. S. and Parcollet, O. and Marianetti, C. A.},
  title   = {Electronic structure calculations with dynamical mean-field theory},
  journal = {Rev. Mod. Phys.},
  volume  = {78},
  pages   = {865},
  year    = {2006},
}

@article{solin1970raman,
  author  = {Solin, S. A. and Ramdas, A. K.},
  title   = {Raman spectrum of diamond},
  journal = {Phys. Rev. B},
  volume  = {1},
  pages   = {1687},
  year    = {1970},
}

@article{huber1964robust,
  author  = {Huber, Peter J.},
  title   = {Robust estimation of a location parameter},
  journal = {Ann. Math. Stat.},
  volume  = {35},
  pages   = {73},
  year    = {1964},
}

\end{document}